\documentclass[sts]{imsart2}
\RequirePackage{amsthm,amsmath,amsfonts,amssymb,xcolor,booktabs,url}
\RequirePackage[numbers]{natbib}

\startlocaldefs
\newcommand{\R}{\mathbb{R}}
\newcommand{\B}{\boldsymbol}
\newcommand{\E}{\mathbb{E}}
\newcommand{\X}{\mathbf{X}}
\newcommand{\x}{\mathbf{x}}
\newcommand{\y}{\mathbf{y}}
\newcommand{\G}{\mathcal{G}}
\newcommand{\Y}{\mathbf{Y}}
\newcommand{\h}{\mathbf{h}}
\DeclareMathOperator*{\argmin}{arg\,min}

\newtheorem{theorem}{Theorem}[section]

\newtheorem{assumption}{Assumption}[section]

\newtheorem{remark}{Remark}[section]%
\newtheorem{definition}{Definition}[section]

\endlocaldefs
\begin{document}

\begin{frontmatter}
\title{Multivariate Distribution-Free Nonparametric Testing: Generalizing Wilcoxon's Tests via Optimal Transport}
\runtitle{Optimal Transport for Nonparametric Testing}

\begin{aug}
\author[A]{\fnms{Zhen}~\snm{Huang}\ead[label=e1]{zh2395@columbia.edu}}
\and
\author[B]{\fnms{Bodhisattva}~\snm{Sen}\ead[label=e2]{bodhi@stat.columbia.edu}}
\address[A]{Zhen Huang is PhD, Department of Statistics, New York, USA\printead[presep={\ }]{e1}.}
\address[B]{Bodhisattva Sen is Professor, Department of Statistics, Columbia University, New York, USA\printead[presep={\ }]{e2}.}
\end{aug}

\begin{abstract}
This paper reviews recent advancements in the application of optimal transport (OT) to multivariate {\it distribution-free nonparametric testing}. Inspired by classical rank-based methods, such as Wilcoxon’s rank-sum and signed-rank tests, we explore how OT-based ranks and signs generalize these concepts to multivariate settings, while preserving key properties, including distribution-freeness, robustness, and efficiency. Using the framework of asymptotic relative efficiency (ARE), we compare the power of the proposed (generalized Wilcoxon) tests against the Hotelling's $T^2$ test. The ARE lower bounds reveal the Hodges-Lehmann \citep{Hodges1956efficiency} and Chernoff-Savage \citep{chernoff1958} phenomena in the context of multivariate location testing, underscoring the high power and efficiency of the proposed methods. We also demonstrate how OT-based ranks and signs can be seamlessly integrated with more modern techniques, such as kernel methods, to develop universally consistent, distribution-free tests. Additionally, we present novel results on the construction of consistent and distribution-free kernel-based tests for multivariate symmetry, leveraging OT-based ranks and signs.
\end{abstract}

\begin{keyword}
\kwd{asymptotic relative efficiency}
\kwd{kernel methods}
\kwd{multivariate signs and ranks}
\kwd{multivariate Wilcoxon's rank-sum test}
\kwd{multivariate Wilcoxon's signed-rank test}
\kwd{two-sample problem}
\kwd{testing for symmetry}
\end{keyword}

\end{frontmatter}
\section{Introduction}
Nonparametric testing is a class of statistical methods that do not rely on strong parametric assumptions. Frank Wilcoxon's 1945 paper \citep{wilcoxon1945individual} was a significant breakthrough in the history of (classical) nonparametric statistics as it introduced two widely used nonparametric tests: the Wilcoxon's {\it rank-sum test} and the Wilcoxon's {\it signed-rank test}. These two tests utilize the ranks of the (univariate) data to test whether two samples come from the same distribution, and whether one sample has center 0, respectively. Moreover, both these rank-based tests are {\it distribution-free}, i.e., the distribution of the test statistics, under the null hypothesis, does not depend on the data generating distribution --- a consequence of the fact that for exchangeable data, the vector of ranks follows the uniform distribution over all $n!$ permutations of $\{1,\ldots,n\}$. 

Apart from their simplicity, robustness and distribution-freeness, a striking feature of these tests that makes them a strong alternative to the usual two-sample $t$-test and paired $t$-test are the efficiency results of Hodges-Lehmann \citep{Hodges1956efficiency} and Chernoff-Savage \citep{chernoff1958}. These results state that Wilcoxon's tests (with Gaussian score) are never worse than the $t$-tests in the sense of {\it asymptotic relative efficiency} (ARE), against location shift alternatives \citep{chernoff1958}. 

Since Wilcoxon’s seminal work~\citep{wilcoxon1945individual}, the use of signs and ranks to develop distribution-free tests has become widespread in one-dimensional nonparametric statistics~\citep{Lehmann75,hajek1999represent,Hettmansperger2011robust}. These methods have been extensively applied in various fields, including time series analysis~\citep{Hallin1992rank,Hallin1988ts,Hirukawa2010rank}, survival analysis~\citep{Gehan1965censor,Prentice1978rank}, finance~\citep{belaire2005variance}, social science~\citep{cliff2014ordinal} and genomics~\citep{Schimek2012rank}. 

In multi-dimensional Euclidean space, the absence of a canonical ordering poses a long-standing challenge in defining multivariate generalizations of these tests which preserve the desirable statistical properties from the one-dimensional case. Since these tests rely on ranks, a more fundamental question that arises is: ``How can one define multivariate ranks that lead to distribution-free testing procedures?". Existing extensions of multivariate ranks --- such as component-wise ranks~\cite{Bickel1965competitors}, spatial ranks~\cite{Chaud96, Kol97, marden1999multivariate}, depth-based ranks~\cite{Liu90, liu1993, Zuo2000}, and Mahalanobis ranks~\cite{Hallin2004, Davy2006} (see \cite{Serfling2010Equivariance} and \cite{hallin2017distribution} for recent surveys) --- and the corresponding rank-based tests do not achieve exact finite-sample distribution-freeness.

A major breakthrough in this regard was recently made in Hallin et al.\ \citep{hallin2017distribution}, where the authors introduced a notion of multivariate ranks based on the theory of optimal transport (OT) (also see~\citep{Chernozhukov2017}). The methodology was later extended by \cite{Nabarun2021rank,huang2023multivariate}. The class of OT-based signs and ranks retains many of the desirable properties of their one-dimensional counterparts. Notably, they are distribution-free in finite samples, and can be used to design distribution-free multivariate analogues of Wilcoxon's tests. These proposed tests~\cite{deb2021efficiency,huang2023multivariate} also exhibit the Hodges-Lehmann and Chernoff-Savage phenomena, ensuring that their power (in the framework of ARE) is at least as high as the two-sample and one-sample Hotelling's $T^2$ tests (the multivariate analogues of the $t$-tests) against many subclasses of location shift alternatives. 

These developments have sparked a surge of research in this area. OT-based signs and ranks have recently been applied to various statistical problems, including the two-sample problem~\citep{boeckel2018multivariate, Nabarun2021rank, deb2021efficiency, ghosal2022multivariate}, independence testing~\citep{Nabarun2021rank, Shi2022indep, ghosal2022multivariate, shi2020rate, shi2023semiparametrically}, and testing for multivariate symmetry~\citep{huang2023multivariate}. They have also been utilized in multiple-output linear regression and MANOVA~\citep{hallin2023fully}, multiple-output quantile regression~\citep{del2024nonparametric}, R-estimation for VARMA models~\citep{Hallin2022VARMA}, and robust testing procedures for VAR models~\citep{Hallin2023VAR}. For a recent review on the applications of OT in statistical problems, see~\cite{Hallin2022review}.

In this paper, we provide a brief review of OT-based (multivariate) signs and ranks and associated statistical methodologies. As illustrative applications, we focus on extending Wilcoxon’s distribution-free tests to the multivariate setting. Using the framework of ARE, we compare the power of the proposed (generalized Wilcoxon) tests against the Hotelling's $T^2$ test. The ARE lower bounds reveal the Hodges-Lehmann \citep{Hodges1956efficiency} and Chernoff-Savage \citep{chernoff1958} phenomena in the context of multivariate location testing, underscoring the high power and efficiency of the proposed methods. We further develop universally consistent two-sample and one-sample tests by integrating OT-based ideas with kernel methods~\cite{Gretton12, sriperumbudur2010}. 

We also discuss a general framework for testing {\it multivariate symmetry} using OT-based (multivariate) {\it signs} and {\it signed-ranks}~\cite{huang2023multivariate}. Our presentation includes novel results (see Theorems~\ref{thm:equiv_cond},~\ref{thm:consistency} and~\ref{thm:recentered_CLT}) not previously published --- specifically, in Sections~\ref{sec:one-sample-kernel-OT} and~\ref{sec:recentered_CLT}, we introduce and study a novel kernel method for testing multivariate symmetry, based on maximum mean discrepancy~\cite{Gretton12} and the OT-based signed-ranks. To the best of our knowledge, this is the only test for multivariate symmetry that satisfies three key properties: (i) distribution-freeness, (ii) universal consistency, and (iii) applicability to a broad class of multivariate symmetries. 

Our approach provides a unified framework for constructing distribution-free tests using OT-based ranks that extend beyond one- and two-sample testing. More broadly, it offers a general mechanism for transforming any test statistic into a distribution-free version by substituting the original observations with OT-based ranks (see Remark~\ref{rem:Framework}).

The paper is organized as follows: Section~\ref{sec:OTprelim} introduces the fundamental concepts of OT. In Section~\ref{sec:OT-multi-ranks}, we review OT-based ranks and demonstrate their application in extending Wilcoxon’s rank-sum test, as well as in designing a universally consistent test for the equality of two distributions. Section~\ref{sec:multi-signed-rank-OT} extends this framework to define OT-based (multivariate) signs and signed-ranks and develops the generalized multivariate Wilcoxon's signed-rank test~\cite{huang2023multivariate}. Our novel methodology and results on testing multivariate symmetry are presented in Sections~\ref{sec:one-sample-kernel-OT} and~\ref{sec:recentered_CLT}. Further applications of OT-based signs and ranks are discussed in Section~\ref{sec:review-further}. Some open problems in this area of distribution-free nonparametric testing are discussed in Section~\ref{sec:Open}. All proofs and simulation results are provided in Appendices~\ref{sec:AppendixProofs} and~\ref{sec:OT-MMD-simu}.


\section{Introduction to Optimal Transport}\label{sec:OTprelim}
{\it Optimal transport} (OT), or ``measure transportation'', is the problem of finding a ``nice'' function ${F}:\R^p\to\R^p$ that transports a given measure $\mu$ on $\R^p$ to another measure $\nu$ on $\R^p$. More specifically, we say ${F}$ pushes $\mu$ to $\nu$, written as ${F}\# \mu = \nu$, if for $\X\sim \mu$, we have ${F}(\X)\sim \nu$. In the 1781 pioneering work of Gaspard Monge \citep{monge1781memoire}, the following problem ({Monge's problem}) was introduced:
\begin{equation}\label{eq:Monge}\inf_{F} \int c(\x , {F}(\x) ){\rm d}\mu(\x),\quad {\rm subject\ to\ } \;\;{F}\#\mu=\nu.
\end{equation}
Here $c(\cdot,\cdot)$ is a cost function. Monge initially worked with the cost function $c(\x,\x') = \|\x-\x'\|$, for $\x,\x'\in\R^p$. A useful and popular cost function is the squared loss $c(\x,\x')=\|\x-\x'\|^2$.
The minimizer of~\eqref{eq:Monge}, if it exists, is referred to as an {\it optimal transport map}.

The main challenge in Monge's problem is that a transport map may not always exist, and the optimization problem is inherently non-convex, making it difficult to solve. To address these limitations, Leonid Kantorovich \citep{Kantorovitch1942} introduced a relaxation of \eqref{eq:Monge} in the 1940s. Let $\Gamma(\mu,\nu)$ denote the set of all joint (Borel) distributions $(\mathbf{X},\mathbf{Y})$ with marginals $\mathbf{X}\sim \mu$ and $\mathbf{Y}\sim \nu$. The Kantorovich problem then seeks to solve
\begin{equation}\label{eq:Kantorovich}
\inf_{\pi \in \Gamma(\mu,\nu)} \int c(\x,\y){\rm d}\pi(\x,\y).
\end{equation}
A key advantage of Kantorovich's formulation is that the optimization problem~\eqref{eq:Kantorovich} is  convex --- and, in fact, linear --- ensuring the existence of a solution under mild conditions, even in cases where no deterministic transport map satisfies Monge’s original formulation.

A fundamental result in this field is Brenier's theorem \citep{Brenier1991}, which states that if $\mu$ is a Lebesgue absolutely continuous distribution on $\R^p$, and both $\mu$ and $\nu$ have finite second moments, and the cost function is given by the squared Euclidean distance $c(\x,\x')=\|\x-\x'\|^2$, then the OT problem~\eqref{eq:Monge} admits a 
$\mu$-a.e.~unique solution. Moreover, this solution is given by the gradient of a convex function. Additionally, under these conditions, Kantorovich's relaxation~\eqref{eq:Kantorovich} is tight, meaning that it has the same optimal value as Monge's problem. Further, the solution to Kantorovich's problem is given by the optimal coupling $\pi = (\mathrm{Id}, F)\#\mu$,\footnote{In this notation, $(\mathrm{Id}, F)\#\mu$ represents the pushforward measure of $\mu$ under the mapping $\x \mapsto (\x,F(\x))$.} where $F$ is the $\mu$-a.e.~unique OT map solving~\eqref{eq:Monge}.  

A significant extension of Brenier’s result was later provided by McCann~\citep{McCann1995}, who adopted a geometric approach and eliminated all moment assumptions. Specifically, if $\mu$ is absolutely continuous on $\R^p$, there exists a $\mu$-a.e.~unique transport map $F$, expressed as the gradient of a convex function, such that $F\#\mu=\nu$. Moreover, if $\mu$ and $\nu$ have finite second moments and $c(\x,\x')=\|\x-\x'\|^2$ (so that problems~\eqref{eq:Monge} and~\eqref{eq:Kantorovich} are well-defined), then $F$ also serves as the OT map in Monge's problem \eqref{eq:Monge}. For a comprehensive treatment of OT, see \cite{villani2009OT, Villani03}.

\section{Multivariate Ranks via OT}\label{sec:OT-multi-ranks}
There are various motivations for the definition of OT-based ranks in the literature \citep{Chernozhukov2017,hallin2017distribution,Nabarun2021rank}. Here we present a simple and natural motivation as given by \cite{Nabarun2021rank}. 

Denote by $\mathcal{P}_{\rm ac}(\mathbb{R}^p)$ the class of all Borel probability measures on $\R^p$ with a Lebesgue density.
Suppose that we are given a collection of $n$ (distinct) i.i.d.\ random variables $X_1,\ldots,X_n$ on $\R$ having a distribution $\mu\in \mathcal{P}_{\rm ac}(\mathbb{R}^p)$.
Let $\mathcal{S}_n$ be the set of all permutations of $\{1,\ldots,n\}$. One can check (by using the rearrangement inequality) that the one-dimensional ranks can be obtained by solving the following problem: 
{\small
$$\hat{\sigma}:=\argmin_{\sigma\in \mathcal{S}_n}\frac{1}{n} \sum_{i=1}^n \Big| X_{\sigma(i)}-\frac{i}{n} \Big|^2,\;\; {\rm and}\quad {\rm rank}(X_i)=\hat{\sigma}^{-1}(i).$$
}
In order to minimize the above quantity, $\sigma$ has to be a permutation that ranks $\{X_1,\ldots,X_n\}$ from the smallest to the largest.

Using the language of OT, the rank map $\hat{R}_n$ assigns $X_1,\ldots,X_n$ to elements of the set $\{\frac{1}{n},\frac{2}{n},...,\frac{n}{n}\}$ by solving the OT problem
\begin{equation}\label{eq:OT_1dim}\hat{R}_n := \argmin_{T:T\# \mu_n=\nu_n}\sum_{i=1}^n |X_i-T(X_i)|^2,\end{equation}
where $\mu_n :=\frac{1}{n}\sum_{i=1}^n \delta_{X_i}$ is the empirical distribution of the observed data (here $\delta_a$ denotes the Dirac delta measure at $a$) and $\nu_n := \sum_{i=1}^n \delta_{i/n}$ is the empirical distribution of $\{\frac{1}{n},\frac{2}{n},\ldots,\frac{n}{n}\}$ --- a natural discretization of Unif$([0,1])$, the uniform distribution on $[0,1]$. 

The OT problem \eqref{eq:OT_1dim} can be easily generalized to higher dimensions.
Suppose $\X_1,\ldots,\X_n$ are i.i.d.\ on $\R^p$ following a Lebesgue absolutely continuous distribution $\mu$. Let $\mu_n :=\frac{1}{n}\sum_{i=1}^n \delta_{\X_i}$, and let $\nu_n$ be the empirical distribution over a known set of (distinct) candidate rank vectors $\{\h_1,\ldots,\h_n\}\subset\R^p$; $\h_1,\ldots,\h_n $ are often chosen such that they provide a discretization of some ``reference distribution'' $\nu$
(some choices of reference distributions are discussed in~\cite[Section 3.3]{deb2021efficiency} and \cite[Remark 3.3]{huang2023multivariate}).
The {\it multivariate sample rank map} $\hat{R}_n:\{\X_1,\ldots,\X_n\} \to \{h_1,\ldots,\h_n\}$ can now be defined as in \eqref{eq:OT_1dim}:
\begin{equation}\label{eq:OT_highdim}
\hat{R}_n := \argmin_{T:T\# \mu_n=\nu_n}\sum_{i=1}^n \|\X_i-T(\X_i)\|^2.\end{equation}
Under the assumption that $\nu_n$ converges weakly to some distribution $\nu$,
one can expect that the sample rank map converges to a {\it population rank map} $R$ defined as:
$${R} := \argmin_{T:T\# \mu = \nu} \E \|\X-T(\X)\|^2,\quad \X\sim \mu.$$
In the one-dimension case, if $\nu\sim {\rm Unif}[0,1]$, the population rank map $R(\cdot)$ is nothing but the cumulative distribution function (c.d.f.) of $\mu$.
In higher dimensions, this mapping is sometimes referred to as the multivariate Brenier c.d.f.~\cite{boeckel2018multivariate}.

\cite[Theorem 2.3]{boeckel2018multivariate} shows that when the population rank maps $R$ is continuous and $\mu$, $\nu$ are compactly supported, {\it uniform sup-norm convergence} holds almost surely (a.s.), i.e., $\sup_{\x\in\R^p}\|\hat{R}_n (\x) - R(\x)\| \to 0$ a.s.~as $n \to \infty$. 
In \cite{hallin2017distribution}, where $\nu$ is chosen as the {\it spherical uniform distribution},\footnote{The spherical uniform distribution is the law of the product of a uniformly distributed vector over the unit sphere in $\R^p$ and an independent uniform variable on $[0,1]$.} the population and sample rank map has an equivariance\footnote{That is, for an orthogonal matrix $\mathbf{O} \in \R^{p \times p}$ and $\mathbf{b}\in\R^p$, the rank map of $\mathbf{O}\X +\mathbf{b}$ to $\nu$, denoted by $R^{\mathbf{O}\X +\mathbf{b}}(\cdot)$, satisfies $R^{\mathbf{O}\X +\mathbf{b}}(\mathbf{O}\x +\mathbf{b})=\mathbf{O}R^\X (\x)$ for all $\x \in \R^p$, where $R^\X(\cdot)$ is the rank map of $\X$ to $\nu$. The sample rank map satisfies a property of the same form, provided that the reference grid used for $\{\mathbf{O}\X_i +\mathbf{b}\}_{i=1}^n$ is $\mathbf{O}\mathfrak{G}_n$, where $\mathfrak{G}_n$ is the reference grid used by $\{\X_i\}_{i=1}^n$. In fact, such equivariance properties hold for any other spherically symmetric reference distributions $\nu$.} property with respect to orthogonal transformations \citep[Proposition 2.2]{hallin2023fully}, and a Glivenko-Cantelli type result holds~\citep[Proposition 2.4]{hallin2017distribution}: $\max_{1\leq i\leq n} \| R(\X_i)-\hat{R}_n(\X_i)\|\to 0$ a.s.~as $n \to \infty$. For a general $\nu$, \cite[Theorem 2.1]{Nabarun2021rank} showed the almost sure $L^2$ convergence of $\hat{R}_n$ to $R$:
$$\frac{1}{n}\sum_{i=1}^n \|\hat{R}_n(\X_i) - R(\X_i)\|\overset{a.s.}{\longrightarrow} 0.$$

The OT-based multivariate ranks defined in \eqref{eq:OT_highdim} exhibit several properties analogous to their classical one-dimensional counterparts, such as {\it maximum ancillarity} \cite{hallin2017distribution}, {independence from the order statistics} \citep{hallin2017distribution,Nabarun2021rank}, and {\it distribution-freeness} \citep{hallin2017distribution,Nabarun2021rank}. The distribution-freeness property states that the rank vector $(R(\X_1),\ldots,R(\X_n))$ is uniformly distributed over all $n!$ permutations of the grid $\{\h_1,\ldots,\h_n\}$, regardless of the underlying data distribution $\mu$. 
This fundamental property enables the development of distribution-free testing procedures based on OT ranks, which have been applied in various statistical problems, including: two-sample testing~\citep{boeckel2018multivariate,Nabarun2021rank,deb2021efficiency}, independence testing \citep{Nabarun2021rank,Shi2022indep,
Mordant2022dep,shi2023semiparametrically}, multivariate linear models~\citep{Hallin2022VARMA,hallin2023fully}, directional data analysis~\citep{Hallin2024directional}, among others.

This distribution-free property enables a simulation-based approach for obtaining critical values for hypothesis tests in finite samples. Suppose that a test statistic $T_n \equiv T_n(R(\X_1),\ldots,$ $R(\X_n))$ is constricted using the OT ranks. To approximate its distribution under the null hypothesis, one can first generate random permutations $\sigma_1,\ldots,\sigma_B$ of the index set $\{1,\ldots, n\}$. The observed test statistic value is then compared against the empirical distribution of $\{T_n(\mathbf{h}_{\sigma_i(1)},\ldots, \mathbf{h}_{\sigma_i(n)})\}_{i=1}^B$
allowing for the computation of a
p-value.

\subsection{Multivariate Wilcoxon's Rank-Sum Test}\label{sec:GWRS}
In this subsection, we discuss the {\it multivariate Wilcoxon's rank-sum test} introduced in \cite{deb2021efficiency}. Suppose we have two independent samples $\{\X_i\}_{i=1}^m \subset \R^p$ and $\{\Y_j\}_{j=1}^n \subset \R^p$, drawn from distributions $\mu_1$ and $\mu_2$, respectively. We are interested in testing the hypothesis:
\begin{equation}\label{eq:2-Sample}
{\rm H}_0:\mu_1=\mu_2 \qquad \mbox{against} \qquad {\rm H}_1:\mu_1\neq \mu_2.
\end{equation}
Similar to its classical one-dimensional counterpart, the multivariate Wilcoxon's rank-sum statistic first computes the sample rank map $\hat{R}_{m,n}$ on the pooled sample, i.e., $\hat{R}_{m,n}: \{\X_1,\ldots,\X_m,\Y_1,\ldots,\Y_n\}\to \{\h_1,\ldots,\h_{m+n}\}$. The test statistic, as proposed in \cite{deb2021efficiency}, is given by:
$$T_{m,n}^\nu := \frac{mn}{m+n} \Big\| \frac{1}{m} \sum_{i=1}^m \hat{R}_{m,n}(\X_i) - \frac{1}{n} \sum_{i=1}^n \hat{R}_{m,n}(\Y_i) \Big\|^2.$$
A more general version of the test statistic  incorporates a score function $J(\cdot):\R^p\to\R^p$, which is injective and Lebesgue a.e.\ continuous, yielding the statistics~\cite{deb2021efficiency}:
$$T_{m,n}^{\nu,J} := \frac{mn}{m+n} (\mathbf{\Delta}_{m,n}^{\nu,J})^\top \Sigma_{\rm ERD}^{-1} \mathbf{\Delta}_{m,n}^{\nu,J},$$
where
$$\mathbf{\Delta}_{m,n}^{\nu,J} := \frac{1}{m}\sum_{i=1}^m J\left(\hat{R}_{m,n}(\X_i)\right) - \frac{1}{n} \sum_{i=1}^n J\left( \hat{R}_{m,n}(\Y_i) \right).$$
Here, $\Sigma_{\rm ERD}$ denotes the  covariance matrix of the push-forward distribution $J\# \nu$. This distribution, referred to as the effective reference distribution (ERD) in \cite{deb2021efficiency}, can be interpreted as the distribution of the score-transformed ranks under the null hypothesis.

Both $T_{m,n}^\nu$ and $T_{m,n}^{\nu,J}$ are functions of the OT-based ranks, so their distributions under the null (i.e., when $\X_i$'s and $\Y_j$'s come from the same distribution) are known. This yields finite sample distribution-free critical values for testing~\eqref{eq:2-Sample}.
For large sample sizes, a normal approximation can be applied. In particular, $T_{m,n}^{\nu,J}$ converges weakly to a chi-squared distribution with $p$ degrees of freedom, as established in~\citep[Theorem 3.1]{deb2021efficiency}.

While distribution-freeness is a highly desirable property, it alone does not guarantee an effective testing procedure. A test must also exhibit good power properties to be practically useful.
A key reason the classical Wilcoxon rank-sum test is a powerful alternative to the two-sample 
$t$-test lies in the seminal results of Hodges-Lehmann \citep{Hodges1956efficiency} and Chernoff-Savage \citep{chernoff1958}. These results state that, against location shift alternatives: the ARE of Wilcoxon’s rank-sum test with respect to the 
two-sample $t$-test is at least 0.864~\citep{Hodges1956efficiency}, and the ARE of the Gaussian score-transformed Wilcoxon test relative to the  $t$-test is at least 1 \citep{chernoff1958}.
This is particularly remarkable because, in the one-dimensional case, the two-sample 
$t$-test is the uniformly most powerful unbiased (UMPU) test when the data are Gaussian \citep{Lehmann2005hypo}. Interestingly, the multivariate Wilcoxon rank-sum test introduced above exhibits a similar efficiency property when compared to Hotelling's 
$T^2$ test~\citep{Hotelling1931}, which is the multivariate analogue of the 
$t$-test.

The Hotelling's $T^2$ test, though classical and simple, remains one of the most popular and useful multivariate two-sample tests today. Its test statistic is given by
$$T_{m,n} := \frac{mn}{m+n} (\bar{\textbf{X}} - \bar{\textbf{Y}})^\top S_{m,n}^{-1} (\bar{\textbf{X}} - \bar{\textbf{Y}}),$$
where $\bar{\X}:=\frac{1}{m}\sum_{i=1}^m \X_i$ and $\bar{\Y}:=\frac{1}{n}\sum_{i=1}^n \Y_i$ are the sample means, and $S_{m,n}$ is the sample covariance matrix of the pooled data.

The {\it asymptotic relative efficiency} (ARE) of a level-$\alpha$ test $T_1$ relative to another level-$\alpha$ test $T_2$ is, roughly speaking,  the limiting ratio of the sample sizes required for $T_2$ to achieve a given power $\beta\in (\alpha,1)$ compared to $T_1$, where the limit is taken along a sequence of alternatives converging to the null.

For instance, if the ARE of $T_1$ with respect to $T_2$ is 0.9, this intuitively means that $T_2$ requires 10\% fewer observations than $T_1$ to attain the same power $\beta$. A formal definition of ARE can be found at \cite[Definition C.2]{deb2021efficiency} (also see~\citep[Chapter 14]{vanderVaart1998}). Although in general the ARE can depend on the choice of $\alpha$ and $\beta$, in our applications it does not.

Although both Hotelling's $T^2$ test and its distribution-free analogue described above can be used to test the general two-sample hypothesis~\eqref{eq:2-Sample}, they are primarily designed for scenarios where the two distributions differ by a location shift, as the tests are based on sample means. This motivates the study of their power behaviors under location-shift alternatives, which we consider next.

Consider a location shift model, where the sample $\{\X_i\}_{i=1}^m$ is drawn from a distribution with Lebesgue density $f_1(\cdot)$, and the sample $\{\Y_j\}_{j=1}^n$ is drawn from a shifted density $f_2 (\cdot)=f_1(\cdot - \mathbf{\Delta})$. 
To compare the two tests in terms of ARE, we examine a sequence of {\it local alternatives} given by $\mathbf{\Delta} = N^{-1/2}\h$, where $N=m+n$ and $\h\neq \mathbf{0}$. Additionally, we assume that the ratio $m/N$ converges to some constant $\lambda\in (0,1)$ as $N\to\infty$. Theorem 3.4 of \citep{deb2021efficiency} states the following result.
\begin{theorem}[Hodges-Lehmann and Chernoff-Savage phenomena]\label{thm:two-sample-Hodge}
Suppose $f_1(\cdot)$ has independent components.\footnote{A random vector $\X=(X_1,\ldots,X_p)$ is said to have independent components if the random variables $X_1,\ldots,X_p$ are independent.} Under mild regularity assumptions,\footnote{The regularity conditions are: (i) The parametric family is quadratic mean differentiable (QMD), which holds for most standard families of distributions, e.g., exponential families in natural form. (ii) The Fisher information matrix exists and is invertible. In addition, $\nu_n$ is required to converge weakly to $\nu$; see~\cite{deb2021efficiency} for the complete statement of the assumptions.} the following results hold:
\begin{enumerate}
\item If $\nu = \mathcal{N}(\mathbf{0},\mathbf{I}_p)$ and the score function is given by $J(\x)=(\Phi(x_1),\ldots,\Phi(x_d))$, where $\Phi$ is the c.d.f.\ of the standard normal distribution, then
the ARE of the multivariate Wilcoxon rank-sum test relative to Hotelling's $T^2$ test is lower bounded by 0.864.
\item If $\nu=\mathcal{N}(\mathbf{0},\mathbf{I}_p)$ and the score function is the identity map, i.e., $J(\x) = \x$, then the ARE of the multivariate Wilcoxon rank-sum test relative to Hotelling's $T^2$ test is lower bounded by 1.
\end{enumerate}
\end{theorem}

The above two scenarios can be interpreted as multivariate analogs of the celebrated results of Hodges-Lehmann \citep{Hodges1956efficiency} and Chernoff-Savage \citep{chernoff1958}, respectively, in a setting where the components of the multivariate vector are assumed to be independent.

Beyond distributions with independent components, the authors in~\citep{deb2021efficiency} also establish similar lower bounds in various other settings, including the {\it elliptically symmetric case}~\citep[Theorem 3.5]{deb2021efficiency}, the {\it blind source separation model}~\citep[Appendix B.2]{deb2021efficiency}, and the {\it Gaussian location problem}~\citep[Proposition 3.3]{deb2021efficiency}. These results extend the Hodges-Lehmann and Chernoff-Savage lower bounds to multivariate settings for specific subclasses of distributions. To the best of our knowledge, no analogous results exist for general location-shift models in the multivariate setting.


\subsection{Two-Sample Tests Based on Rank Kernel MMD}\label{sec:RankMMD}
While the multivariate Wilcoxon's rank-sum test (from Section~\ref{sec:GWRS}) is powerful in location shift models (in terms of ARE against Hotelling's $T^2$ test as seen in the previous section), it is however not consistent against all (fixed) alternatives. When the difference between the two samples does not come from a location shift, the multivariate Wilcoxon's rank-sum test could be powerless.
One approach to gain (universal) consistency while still being distribution-free is to combine the OT-based ranks with kernel method, which will be described below.

A kernel $K(\cdot,\cdot)$ is a {\it positive definite} function, meaning that for any $n\geq 1$, $\x_1,\ldots,\x_n\in\R^p$ and $c_1,\ldots,c_n\in\R$, we have $\sum_{i,j=1}^n c_ic_j K(\x_i,\x_j)\geq 0$.
$K(\cdot,\cdot)$ is called a {\it characteristic} kernel, when the following holds: $\mu_1=\mu_2$ if and only if
$\E_{\X \sim \mu_1}K(\X,\cdot) = \E_{\X \sim \mu_2}K(\X,\cdot)$. Examples of characteristic kernels include 
Gaussian kernel $K(u,v) := \exp(-\sigma\|u - v\|^2)$,
Laplace kernel $K(u,v) = \exp(-\sigma\|u - v\|_1)$ (here $\|\cdot\|_1$ denotes the $L_1$-norm); and
distance kernel $K(u,v):= {2}^{-1}(\lVert u\rVert^{\alpha} +\lVert v\rVert^{\alpha} - \lVert u-v\rVert^{\alpha})$,
for $\alpha \in (0,2)$ \citep{EquivRKHS13}. See \cite{sriperumbudur2008injective,fukumizu2009characteristic,sriperumbudur2010,charac-Bharath,szabo2017characteristic} for sufficient conditions that imply that $K(\cdot,\cdot)$ is characteristic. Given a characteristic kernel $K(\cdot,\cdot)$, the maximum mean discrepancy (MMD) \citep{Gretton12} is a distance on the space of probability distributions defined as
\begin{equation}\label{eq:MMD}
    \begin{aligned}
&\ {\rm MMD}^2(P,Q) := \E K(\Delta_1^P,\Delta_2^P) \\
& \qquad \qquad + \,\E K(\Delta_1^Q,\Delta_2^Q) - 2\E K(\Delta_1^P,\Delta_1^Q),
\end{aligned}
\end{equation}
where $\Delta_1^P,\Delta_2^P\sim P$, $\Delta_1^Q,\Delta_2^Q\sim Q$ are mutually independent.

When the two samples $\{\X_i\}_{i=1}^m$ and $\{\Y_j\}_{j=1}^n$ follow the same distribution, their score-transformed ranks 
$J(\hat{R}_{m,n}(\X_i))$ and $J(\hat{R}_{m,n}(\Y_j))$ will also share the same distribution. Motivated by this fact, \cite[Section 4]{deb2021efficiency} proposed to use the empirical MMD between the score transformed ranks $\{J(\hat{R}_{m,n}(\X_i))\}_{i=1}^m$ and $\{J(\hat{R}_{m,n}(\Y_j)\}_{j=1}^n$ as the test statistic:

\begin{equation*}
    \gamma_{m,n}^{\nu, J} := \frac{mn}{m+n} \left[ w_{m,n}^{(1)} + w_{m,n}^{(2)} - b_{m,n} \right],
\end{equation*}
where
{\small
$$
\begin{aligned}
    w_{m,n}^{(1)} &:= \frac{1}{m(m-1)} \sum_{1 \leq i \neq j \leq m} K \left( J(\hat{R}_{m,n}(X_i)), J(\hat{R}_{m,n}(X_j)) \right), \\
    w_{m,n}^{(2)} &:= \frac{1}{n(n-1)} \sum_{1 \leq i \neq j \leq n} K\left( J(\hat{R}_{m,n}(Y_i)), J(\Hat{R}_{m,n}(Y_j)) \right), \\
    b_{m,n} &:= \frac{2}{mn} \sum_{1 \leq i \leq m} \sum_{1 \leq j \leq n} K\left( J(\hat{R}_{m,n}(X_i)), J(\hat{R}_{m,n}(Y_j)) \right).
\end{aligned}
$$
}

As a function of the OT ranks, $\gamma_{m,n}^{\nu, J}$ is exactly distribution-free for all $m,n\geq 1$, from which the critical value for a level $\alpha$ test can be obtained (for example, via the simulation approach mentioned before Section~\ref{sec:GWRS}).
Apart from distribution-freeness, it also retains the universal consistency property of the classical MMD two-sample test:
\begin{theorem}[Universal consistency; Theorem 4.1 of \cite{deb2021efficiency}]
Let the score function $J(\cdot)$ be injective. Under mild assumptions on the reference distribution and the kernel, and $\frac{m}{m+n}\to \lambda\in(0,1)$, the two-sample test that rejects 
${\rm H}_0:\mu_1=\mu_2$ for large values of $\gamma_{m,n}^{\nu, J}$ is consistent against all (fixed) alternatives where $\mu_1\neq \mu_2$.
\end{theorem}

Under ${\rm H}_0$, \cite[Theorem 4.2]{deb2021efficiency} shows that $\gamma_{m,n}^{\nu, J}$ converges in distribution to an infinite sum of weighted centered chi-squares. \cite[Theorem 4.3]{deb2021efficiency} also shows that the test based on $\gamma_{m,n}^{\nu, J}$ has nontrivial power against $O(1/\sqrt{m+n})$ alternatives, which implies that the test is not only consistent and distribution-free, but also has non-trivial Pitman efficiency.


\begin{remark}[A framework for constructing distribution-free tests]\label{rem:Framework}
OT-based ranks provide a general mechanism for transforming any test statistic into a distribution-free version by replacing the original observations with their pooled OT-based ranks. This approach, analogous to how we obtained the (generalized) Wilcoxon rank-sum test in Section~\ref{sec:GWRS} and the rank kernel MMD test in Section~\ref{sec:RankMMD}, yields distribution-free and robust variants of the original test statistics~\citep{Nabarun2021rank}.

Moreover, this principle extends beyond two-sample testing; see Section~\ref{sec:review-further} for additional applications, including tests for independence between two random vectors. In the context of independence testing, marginal OT-based ranks are used to construct distribution-free test statistics, following the same principle that underlies the derivation of Spearman's rank correlation~\citep{spearman1904proof} from Pearson's correlation coefficient.
\end{remark}

\section{Multivariate Signed-Ranks via OT}\label{sec:multi-signed-rank-OT}
We have seen that the OT-based multivariate Wilcoxon's rank-sum test (in Section~\ref{sec:GWRS}) has ARE lower bounds w.r.t.\ two-sample Hotelling's $T^2$ test, mimicking the relationship between the classical Wilcoxon's rank-sum test and the two-sample $t$-test. As the Wilcoxon's signed-rank test also has ARE guarantees relative to the one-sample $t$-test \citep{Hodges1956efficiency,chernoff1958}, a natural question to ask is whether there exists an OT-based multivariate signed-rank test with similar ARE lower bounds relative to the one-sample Hotelling's $T^2$ test? A positive answer was provided in \cite{huang2023multivariate}, and will be described below.

Wilcoxon's signed-rank test is based on the following concepts. In the classical (one-dimensional) setting, given a set of (distinct) observations $X_1,\ldots,X_n \in \R$, the {\it absolute rank} of $X_i$ is the rank of $|X_i|$ among $|X_1|,\ldots,|X_n|$. The {\it sign} of $X_i$ is $+1$ if $X_i \ge 0$ and is $-1$ otherwise. Further, the {\it signed-rank} of $X_i$ is just the product of it's absolute rank and sign. Similar to Section~\ref{sec:OT-multi-ranks}, it can be checked that the signs and absolute ranks can be obtained via solving the optimization problem:
\begin{equation}\label{eq:1-D-Sign-Rank}
\min \left\{\sum_{i=1}^n | q_i {X}_{\sigma(i)}-i/n |^2:q_i \in \pm 1, \sigma\in \mathcal{S}_n\right\}.
\end{equation}
Suppose $\hat{\sigma}$ and $\{\hat{q}_i\}_{i=1}^n$ are the optimizers of the above problem.
Then: (i) the sign of $X_{\hat{\sigma}(i)}$ is $\hat{q}_i$, (ii) $\hat{q}_i X_{\hat{\sigma}(i)}=|X_{\hat{\sigma}(i)}|$, and (iii) the absolute rank of $X_{\hat{\sigma}(i)}$ is $i$. The above optimization problem will motivate their multivariate analogues; see~\eqref{eq:OTproblem} below.

Observe that in one dimension, the signs and absolute ranks are not immediately distribution-free. Their distribution-freeness requires the additional assumption of {\it symmetry}, i.e., $X$ is equal in distribution to $-X$ (denoted by $X\overset{d}{=}-X$).
If $X_1,\ldots,X_n$ are i.i.d.\ from a one-dimensional absolutely continuous distribution that is symmetric, then the following distribution-free properties of signs and absolute ranks hold: $\{{\rm sign}(X_i)\}_{i=1}^n$ are i.i.d.\ uniform over $\{-1,1\}$, and the ranks of $|X_1|,\ldots,|X_n|$ are uniform over all permutations of $1,\ldots,n$. Moreover, the signs and the absolute ranks are independent.

While the meaning of symmetry in one dimension is quite unambiguous, in the multivariate setting, there exist various notions of symmetry. The most prominent ones include: (a) {\it central symmetry}, 
(b) {\it sign symmetry}, 
and (c) {\it spherical symmetry}; 
a thorough review of the topic can be found in \cite{Serfling2014}.
\cite{huang2023multivariate} consolidated these notions of symmetry and defined OT-based multivariate signs and signed-ranks under a general notion of {\it $\G$-symmetry}; see below. The hypothesis of $\G$-symmetry \citep{beran1997multivariate,huang2023multivariate} is formally defined via a compact subgroup $\G$ of the orthogonal group ${\rm  O}(p)$:
\begin{equation}\label{eq:hypo}
{\rm H}_{\G,0}\ (\G \text{-symmetry}): {\bf X}  \stackrel{d}{=}  Q  {\bf X}{\rm\ \ for\ all\ \ } Q\in \G.
\end{equation}
The prominent examples of symmetry mentioned above correspond to the following groups:
\begin{enumerate}
\item[(a)] {\it Central symmetry}: Here $\G = \{\mathbf{I}, -\mathbf{I}\}$, where $\mathbf{I}\equiv \mathbf{I}_p$ is the identity matrix of order $p$. Central symmetry (or ``reflective'' or ``diagonal'' or ``simple'' or ``antipodal'' symmetry \citep{PuriSen1967,Blough1989PP}) 
arises naturally from paired data 
(see e.g., \cite[Section 5]{Gibbons2003nonpara}):
Given a sample of $n$ pairs $\{(\mathbf{Y}_i,\mathbf{Z}_i)\}_{i=1}^n$ (e.g., treatment-control pairs),
it is often of interest to study the differences
$\{\mathbf{X}_i:=\mathbf{Y}_i-\mathbf{Z}_i\}_{i=1}^n$.
Under the null hypothesis that $\mathbf{Y}_i$ and $\mathbf{Z}_i$ are exchangeable (e.g., no treatment effects), $\mathbf{X}_i$ will have a centrally symmetric distribution \citep{Lehmann75, Gibbons2003nonpara}.

\item[(b)] {\it Sign symmetry}: Here $\G = \{Q \in \R^{p \times p}: Q$ is a diagonal orthogonal matrix$\}$; thus each diagonal entry of $Q$ is $\pm 1$. Sign symmetry is also called {\it orthant symmetry} or {\it marginal symmetry} \citep{efron1969,Pinelis1994T2,Oja2010nonp}. 

\item[(c)] {\it Spherical symmetry}: Here $\G = {\rm O}(p)$. Spherically symmetric distributions are natural generalizations of the Gaussian distribution and have attracted substantial interest~\citep{Fang1990symm,alb2020sph,Jammalamadaka1987linear,Berk1989opt,Fourdrinier1995sph,Li2005contour,Zeng2010central}.
\end{enumerate}

Suppose $\mathbf{X}_1,\ldots,\mathbf{X}_n$ are i.i.d.\ from $\mu\in \mathcal{P}_{\rm ac}(\mathbb{R}^p)$. 
As in Section~\ref{sec:OT-multi-ranks}, let $\{\mathbf{h}_j\}_{j=1}^n \subset\R^p$ be some (distinct) known vectors that provide a discretization of a reference measure $\nu$. The OT-based signs and absolute ranks \citep{huang2023multivariate} are defined below mimicking their one-dimensional counterparts obtained via~\eqref{eq:1-D-Sign-Rank}; they are distribution-free under ${\rm H}_{\G,0}$ where $\X$ is $\G$-symmetric~\cite{huang2023multivariate}.


\begin{definition}[OT-based signs, absolute ranks, and signed-ranks; sample version]
Let $\mathcal{S}_n$ be the set of all $n!$ permutations of $\{1,\ldots ,n\}$. Consider the optimization problem
\begin{equation}\label{eq:OTproblem}
\min \left\{\sum_{i=1}^n \lVert Q_i^\top \mathbf{X}_{\sigma(i)}-\mathbf{h}_i\rVert^2:Q_i \in \G, \sigma\in \mathcal{S}_n\right\}.
\end{equation}
Let $(\{\hat{Q}_i\}_{i=1}^n,\hat{\sigma})$ be the minimizer of \eqref{eq:OTproblem}. 
Define the {\it generalized sign} $S_n:\{\mathbf{X}_1,\ldots,\mathbf{X}_n\}\to \G$ and the {\it generalized absolute rank} $R_n:\{\mathbf{X}_1,\ldots,\mathbf{X}_n\}\to\{\mathbf{h}_1,\ldots,\mathbf{h}_n\}$ as:
\begin{equation*}\label{eq:Sign-Rank}
S_n(\mathbf{X}_i) := \hat{Q}_{\hat{\sigma}^{-1}(i)},\quad {\rm and}\quad R_n(\mathbf{X}_i) := \mathbf{h}_{\hat{\sigma}^{-1}(i)},
\end{equation*}
for $i = 1,\ldots, n$.
The {\it generalized signed-rank} of the $i$-th observation $\mathbf{X}_i$ is then defined as $S_n(\mathbf{X}_i)R_n(\mathbf{X}_i) $.
\end{definition}

Thus, the OT-based signs are $p\times p$ orthogonal matrices belonging to the group $\G$, and the OT-based signed-rank of $\X_i$ is the vector in the orbit\footnote{The {\it orbit} of $\mathbf{h}$ under the action of $\G$ is the set of all elements that can be reached by applying elements of $\G$ to $\mathbf{h}$, i.e., $\{Q\mathbf{h}:Q\in\G\}$.} of the absolute rank $R_n(\X_i)$ that is closest to $\X_i$.

In fact, the OT-based signs, absolute ranks, and signed-ranks defined in \eqref{eq:OTproblem} reduce to their classical definitions when $p=1$, $\G=\{-1,1\}$, and $h_j=\frac{j}{n}$ \citep[Remark 2.1]{huang2023multivariate}. The definition of OT-based signs and signed-ranks can also be thought of as an extension to the OT-based ranks (defined in Section~\ref{sec:OT-multi-ranks}).
In particular, if $\G=\{\mathbf{I}_p\}$ is the trivial group, then the optimization problems~\eqref{eq:OTproblem} and~\eqref{eq:OT_highdim} are exactly the same.

The minimizer of \eqref{eq:OTproblem} is well-defined under mild assumptions. Note that the optimization problem \eqref{eq:OTproblem} is basically the same if we replace $\mathbf{h}_i$ by another vector in the orbit of $\mathbf{h}_i$ --- the minimum of \eqref{eq:OTproblem} remains unchanged. If $\{\mathbf{h}_i\}_{i=1}^n$ is suitably chosen such that no two $\h_i$'s lie on the same orbit of $\G$, then the optimizer $\hat{\sigma}$ (and thus the absolute ranks) is a.s.\ unique \citep[Proposition 2.1]{huang2023multivariate}. $\{\hat{Q}_i\}_{i=1}^n$ (and thus the generalized signs) is also a.s.\ unique if $\G$ induces a {\it free} group action\footnote{Suppose $\{\mathbf{h}_i\}_{i=1}^n\subset B\subset \R^p$. $\G$ acts {\it freely} on $\G B$ if for $\mathbf{y}\in B$ and $Q\in \G$, $Q \mathbf{y}=\mathbf{y}$ implies $Q= \mathbf{I}_p$ (i.e., for any $\mathbf{x}$ in $\G B$, only the identity matrix in $\G$ leaves $\mathbf{x}$ fixed).} (intuitively, a free group action means that given two vectors on the same orbit of $\G$, there is a unique element in $\G$ that moves one to the other). Such a free group action is available for {\it central symmetry} 
and {\it sign symmetry}. 
However, a free group action (and hence the uniqueness of OT-based signs) is not available for {\it spherical symmetry} 
in higher dimensions.\footnote{Take a simple example: Suppose $n=1$, $\mathbf{X}_1=(0,1)^\top$, and $\mathbf{h}_1=(1,0)^\top$. Then the reflection along $y=x$ and a rotation by $\pi/2$ both transport $\mathbf{X}_1$ to $\mathbf{h}_1$ with minimum cost 0. Thus the OT-based sign is not unique.} In such cases when the minimizers $\{\hat{Q}_i\}_{i=1}^n$ are not unique, \cite{huang2023multivariate} proposed to select a minimizer uniformly at random as the OT sign. This selection of signs ensures the signs follow the uniform distribution over $\G$ under the null. Note that even in cases where OT signs are not unique (e.g., when testing spherical symmetry), the OT signed-ranks (that will be used to construct the multivariate Wilcoxon's signed-rank test) are a.s.\ unique.

These generalized notions of signs and absolute ranks defined via OT extend many natural properties of their classical counterparts. For example, the distribution-free properties:
(i) The absolute ranks $(R_n(\mathbf{X}_1),\ldots, R_n(\mathbf{X}_n))$ is independent of any order statistics\footnote{An {\it order statistic} is an un-ordered version of the data $\{\mathbf{X}_1,\ldots,\mathbf{X}_n\}$ \citep{hallin2017distribution}.} of $\{\mathbf{X}_1,\ldots,\mathbf{X}_n\}$, and is uniformly distributed over the set of all $n!$ permutations of $\{\mathbf{h}_i\}_{i=1}^n$; (ii) the absolute ranks $(R_n(\mathbf{X}_1),\ldots,$ $ R_n(\mathbf{X}_n))$ and the generalized signs $(S_n(\mathbf{X}_1),\ldots,S_n(\mathbf{X}_n))$ are independent under ${\rm H}_{\G,0}$; and (iii) $S_n(\mathbf{X}_1),\ldots,S_n(\mathbf{X}_n)$ are i.i.d.\ following the uniform distribution over $\G$, under ${\rm H}_{\G,0}$ \citep[Proposition 2.2]{huang2023multivariate}.

Define cost function
\begin{equation}\label{eq:cost}
c(\mathbf{x},\mathbf{h}):=\min_{Q\in \G}\|Q^\top \mathbf{x}-\mathbf{h}\|^2 , \qquad \mbox{for} \;\; \mathbf{x},\mathbf{h} \in \R^p.
\end{equation}
Then \eqref{eq:OTproblem} reduces to $\min_{\sigma \in \mathcal{S}_n} \sum_{i=1}^n c(\mathbf{X}_{\sigma(i)},\mathbf{h}_i)$ --- the OT problem from $\{\mathbf{X}_i\}_{i=1}^n$ to $\{\mathbf{h}_i\}_{i=1}^n$ under the cost function $c(\cdot,\cdot)$. Therefore, the population OT problem for the absolute ranks can be naturally defined as
\begin{equation}\label{eq:popu_ot_problem}
\inf_{(\mathbf{X},\mathbf{H}):(\mathbf{X},\mathbf{H})\in \Pi(\mu,\nu)} \mathbb{E} \left[c(\mathbf{X},\mathbf{H})\right],
\end{equation}
where $\Pi(\mu,\nu)$ consists of all Borel distributions $(\mathbf{X},\mathbf{H})$ with marginals $\mathbf{X}\sim \mu$ and $\mathbf{H}\sim \nu$.

With the following Assumption~\ref{assump:nu-G}, the solution to \eqref{eq:popu_ot_problem} (i.e., the population absolute rank map) is unique and is given by a deterministic map $R(\x)$ characterized in \cite[Theorem 2.1]{huang2023multivariate} (see Theorem~\ref{thm:population_rank} below). 

\begin{assumption}\label{assump:nu-G}
There exists a Borel set $B\subset\R^p$ with $\nu(B)= 1$ such that for any $\mathbf{x}\in\mathbb{R}^p$, the orbit\footnote{The orbit of an element $\x \in \R^p$ is the set of elements (in $\R^p$) to which $\x$ can be moved by the elements of $\G$.} of $\mathbf{x}$, i.e., $\{Q\mathbf{x}: Q\in \G\}$, intersects $B$ at one point at most.
\end{assumption}
This assumption requires that at most one point should be taken as a rank vector (which belongs to $B$) from any orbit of $\G$. This is because transporting $\x$ to any point in an orbit of $\G$ has the same cost due to the form of the cost function $c(\cdot,\cdot)$ (see \eqref{eq:cost}). So it is not necessary to include two representative points from the same orbit, which may break the uniqueness of the population OT map.\footnote{For example, if $\mu = \nu = \mathcal{N}(\mathbf{0},\mathbf{I})$, then $R(\mathbf{x})= \mathbf{x}$ and $R(\mathbf{x})=- \mathbf{x}$ are both OT maps for \eqref{eq:popu_ot_problem} with 0 loss, if $-\mathbf{I}\in \G$.}

\begin{theorem}[Population absolute rank and signed-rank maps. Theorem 2.1 of \cite{huang2023multivariate}]\label{thm:population_rank}
Suppose Assumption~\ref{assump:nu-G} holds. Denote the distribution of $\mathbf{X}$ by $\mu\in \mathcal{P}_{\rm ac}(\mathbb{R}^p)$. 
Then there exists a $\mu$-a.e.\ unique Borel measurable map $R:\R^p\to\R^p$ such that $(\mathbf{X},R(\mathbf{X}))$ has the unique distribution in $\Pi(\mu,\nu)$ with a $c$-cyclically monotone\footnote{For a function $c(\cdot,\cdot)$ on $\R^p\times\R^p$, a subset $\Gamma\subset \R^p\times \R^p$ is said to be $c$-cyclically monotone if for any $N\in\mathbb{N}$ and any family $(\mathbf{x}_1,\mathbf{y}_1),\ldots,(\mathbf{x}_N,\mathbf{y}_N)$ of points in $\Gamma$, the inequality
$\sum_{i=1}^Nc(\mathbf{x}_i,\mathbf{y}_i)\leq \sum_{i=1}^Nc(\mathbf{x}_i,\mathbf{y}_{i+1})$ holds
with the convention $\mathbf{y}_{N+1}=\mathbf{y}_1$.
} support. Moreover, $R$ has the following properties:
\begin{enumerate}
\item[(i)]
Let $q$ be the quotient map\footnote{Under Assumption~\ref{assump:nu-G}, the quotient map $q:\G B\to B$ is defined by $q(Q\mathbf{x})=\mathbf{x}$ for $\mathbf{x}\in B$, $Q\in \G$.
Assumption~\ref{assump:nu-G} ensures $q$ is well-defined.
} from $\G B$ to $ B$. Then, there exists a lower semicontinuous convex function $\psi:\mathbb{R}^p\to (-\infty,+\infty]$ such that
$$R(\mathbf{x}) = q(\nabla \psi(\mathbf{x})) \quad (\mu\mbox{-a.e.}~\mathbf{x}).$$
\item[(ii)] Let $S\sim {\rm Uniform}(\G)$ be independent of $\mathbf{X}\sim \mu$ and $\mathbf{H}\sim\nu$. Let $\mu_S$ and $\nu_S$ denote the distributions of $S\mathbf{X}$ and $S\mathbf{H}$, respectively. Then
$\nabla \psi(\cdot)$ is the $\mu_S$-a.e.\ unique gradient of convex function that pushes $\mu_S$ to $\nu_S$.
\item[(iii)] $\nabla \psi$ is equivariant under the group action of $\G$, i.e.,
$\nabla \psi(Q\mathbf{x}) = Q\nabla \psi(\mathbf{x})$ for $Q\in \G$ and $\mathbf{x} \in \R^p$ being a differentiable point of $\psi$.
\item[(iv)] $\nabla \psi(\mathbf{X}) \overset{a.s.}{=} S(\mathbf{X},R( \mathbf{X})) R(\mathbf{X})$ with $S(\mathbf{x},\mathbf{h}) := \argmin_{Q\in \G} \|Q^\top\mathbf{x} - \mathbf{h}\|^2$. Hence, $\nabla \psi(\cdot)$ can be viewed as the population signed-rank map.
\item[(v)] If $\mu$ and $\nu$ have finite second moments, then
$({\rm identity}, R)\#\mu$ is the unique solution to the population OT problem \eqref{eq:popu_ot_problem}.
\end{enumerate}
\end{theorem}
Let us discuss the implications of the above results:
(ii) and (iv) state that the population signed-rank map $\nabla \psi(\x)$ is the (unique) gradient of a convex function that maps the symmetrized data distribution $\mu_S$ to the symmetrized reference distribution $\nu_S$; 
(i) characterizes the population absolute rank map, which is $\nabla \psi(\x)$ composed with the quotient map $q(\cdot)$ that brings it to $B$ ($\nu$ is concentrated on $B$);
(iii) illustrates a `nice' property of the population signed-rank map $\nabla \psi(\x)$ --- it respects the symmetry of $\G$. Note that (i)-(iv) do not assume any moment conditions on $\mu$ and $\nu$. They are based on a geometric characterization of the rank map using $c$-cyclical monotonicity. When second moments of $\mu$ and $\nu$ do exist, (v) shows that the population rank map is indeed the OT map in the sense of minimizing \eqref{eq:popu_ot_problem}.

Assuming that $\nu_n = \frac{1}{n} \sum_{i=1}^n \delta_{\mathbf{h}_i}$ converges weakly to the reference distribution $\nu$, \cite[Theorem 2.2]{huang2023multivariate} shows the sample absolute rank map $R_n(\cdot)$ and the sample signed-rank map $S_n(\cdot)R_n(\cdot)$ converge to their population versions $R(\cdot)$ and $\nabla\psi(\cdot)$ respectively. If furthermore $\G$ induces a free group action (so that signs are uniquely defined), then the sample sign map $S_n(\cdot)$ will also converge to a population sign map given by $S(\x)=\argmin_{Q\in\G}\|Q^\top \x -R(\x)\|$.

\subsection{Multivariate Wilcoxon's Signed-Rank Test}
With the notions of OT-based signs and absolute ranks defined above, the multivariate Wilcoxon's signed-rank test statistic can be defined as $\sum_{i=1}^n S_n(\X_i)R_n(\X_i)$. With a score function $J:\R^p\to\R^p$, a slightly more general score transformed version is given by
\begin{equation*}\label{eq:score_Wn}\mathbf{W}_n:= \frac{1}{\sqrt{n}}\sum_{i=1}^n S_n(\mathbf{X}_i)  J(R_n(\mathbf{X}_i)).
\end{equation*}
Note that $\mathbf{W}_n$ is a function of the OT-based signs and ranks, so it is distribution-free under ${\rm H}_{\G,0}$ \citep[Proposition 2.2]{huang2023multivariate}.
The known distribution in finite samples can be used to obtain the critical values for the test based on $\mathbf{W}_n$.

While the multivariate Wilcoxon's signed-rank test has a distribution-free property under ${\rm H}_{\G,0}$ and can be used to test ${\rm H}_{\G,0}$, it is much more useful for the following sub-problem (the {\it one-sample location} problem). 
Suppose $\mathbf{X}_1,\ldots,\mathbf{X}_n$ are i.i.d.\ with Lebesgue density $f(\mathbf{\cdot} -\mathbf{\Delta})$ on $\R^p$, where $f$ is the density of a $\G$-symmetric distribution and $\mathbf{\Delta} \in \R^p$. Consider testing
\begin{equation}\label{eq:location-hypo}
{\rm H}_0: \mathbf{\Delta} = \mathbf{0}_p \qquad\textrm{against}\qquad {\rm H}_1: \mathbf{\Delta} \neq \mathbf{0}_p.
\end{equation}

\cite{huang2023multivariate} suggests rejecting ${\rm H}_0$ for large values of 
\begin{equation*}\label{eq:Wilcox-Test}
\mathbf{W}_n^\top \Sigma_{\rm ERD_1}^{-1}\mathbf{W}_n,
\end{equation*}
where $\Sigma_{\rm ERD_1}$ is the covariance matrix of $SJ(\mathbf{H})$ --- $S\sim {\rm Uniform}(\G)$ is independent of $\mathbf{H}\sim \nu$. The distribution of $SJ(\mathbf{H})$ is referred to as the {\it effective reference distribution} (${\rm ERD}_1$) --- to distinguish it from the ERD in Section~\ref{sec:OT-multi-ranks}, we write is as ${\rm ERD}_1$ (`1' denotes `one-sample').
As $n\to\infty$, $\mathbf{W}_n$ has an asymptotic normal distribution $\mathcal{N}(\mathbf{0}_p,\Sigma_{\rm ERD_1})$ \citep[Theorem 3.1]{huang2023multivariate}, and thus $\mathbf{W}_n^\top \Sigma_{\rm ERD_1}^{-1}\mathbf{W}_n$ asymptotically follows a chi-squared distribution with $p$ degrees of freedom. Critical values for the asymptotic test thus can be obtained easily.
The consistency of the multivariate Wilcoxon's signed-rank test for testing \eqref{eq:location-hypo} is established in \cite[Corollary 3.1]{huang2023multivariate}

It is natural to compare the multivariate Wilcoxon's signed-rank test with one-sample Hotelling's $T^2$ test, whose test statistic is given by 
\begin{equation*}\label{eq:Hotelling-T2}
T^2=n\bar{\mathbf{X}}_n^\top S_n^{-1} \bar{\mathbf{X}}_n,
\end{equation*}
where $S_n := \frac{1}{n-1}\sum_{i=1}^n (\mathbf{X}_i-\bar{\mathbf{X}}_n)(\mathbf{X}_i-\bar{\mathbf{X}}_n)^\top$ is the sample covariance matrix. 

Recall the one-sample location problem \eqref{eq:location-hypo}. Consider the local alternatives:
\begin{equation}\label{eq:local_test}
{\rm H}_0: \mathbf{\Delta} = \mathbf{0}_p \qquad\textrm{versus}\qquad {\rm H}_1: \mathbf{\Delta} = \frac{{\B \xi}_n}{\sqrt{n}},
\end{equation}
where ${\B \xi}_n\to {\B \xi}\in\R^p$ as $n \to \infty$. The ARE of the multivariate Wilcoxon's signed-rank test relative to the one-sample Hotelling's $T^2$ test is lower bounded in various cases \citep[Proposition 3.2, Theorems 3.5, 3.6, 3.7]{huang2023multivariate}. For example, in the independent components cases, we have the following result.

\begin{theorem}[Hodges-Lehmann and Chernoff-Savage phenomena; Theorems 3.5 of \cite{huang2023multivariate}]
Suppose $f(\cdot)$ has independent components. Under mild regularity conditions,\footnote{Similar to Theorem~\ref{thm:two-sample-Hodge}, the regularity conditions are standard quadratic mean differentiability of the parametric family $\{\mathcal{P}_\mathbf{\theta}\}_{\theta\in\R^p}=\{f(\cdot -\theta):\theta\in\R^p\}$ and the invertibility of the Fisher information matrix at the origin $\theta=\mathbf{0}_p$. In addition, $\nu$ is required to converge $\nu$ weakly and in 2nd moment; see \cite{huang2023multivariate} for a detailed description of the assumptions.} we have
\begin{enumerate}
\item If the ${\rm ERD}_1$ is Unif$[-1,1]^d$, with $J(\x)=\x$, then
the ARE of multivariate Wilcoxon's signed-rank test relative to one-sample Hotelling's $T^2$ test is lower bounded by 0.864.
\item If the ${\rm ERD}_1$ is $\mathcal{N}(\mathbf{0},\mathbf{I}_d)$, with $J(\x)=\x$, then the ARE of multivariate Wilcoxon's signed-rank test relative to one-sample Hotelling's $T^2$ test is lower bounded by 1.
\end{enumerate}
\end{theorem}
The above result can again be seen as a multivariate analogue to the Hodges-Lehmann and Chernoff-Savage phenomena for the classical Wilcoxon's signed-rank test. Note that the above result is valid for all possible choices of $\G$ as long as the ${\rm ERD}_1$ is of the specified form.

In fact, if the score function $J(\cdot )$ is suitably chosen, there exists a (score transformed) multivariate Wilcoxon's signed-rank test that is {\it locally asymptotically optimal}, i.e., it has maximum local power (against contiguous alternatives) among all tests with fixed Type I error; see \cite[Theorem 3.9]{huang2023multivariate} for details.

\subsection{Universally Consistent Tests for Multivariate Symmetry}\label{sec:one-sample-kernel-OT}
While the multivariate Wilcoxon's signed-rank test is highly powerful in detecting deviations in location~\eqref{eq:location-hypo} and consistent against a class of alternatives including location shift alternatives~\citep[Theorem 3.2]{huang2023multivariate}, it is however not necessarily consistent or powerful against the general alternative of ${\rm H}_{\G,0}$ in \eqref{eq:hypo}.
Note that the multivariate Wilcoxon's signed-rank test rejects the null when the sum of signed-ranks is far from the origin, therefore it requires the signed-ranks to have mean $\mathbf{0}$ under the null to be powerful, formulated by the assumption $-\mathbf{I}_p\in \G$ in its consistency theorem \cite[Corollary 3.1]{huang2023multivariate}.
This excludes symmetries such as exchangeability, reflection, and zonal symmetry:
\begin{enumerate}
\item {\it Exchangeability} \citep{bell1969bivariate,Hollander1971sym,Snijders1981sym,modarres2008tests,Oja2010nonp,EG2016,kalina2022testing}. Here, $\G$ consists of all permutation matrices of order $p$. Exchangeability is an important concept in various statistical models \citep{Vovk2005algo,Oja2010nonp} and can be related to treatment-control pairs, as bivariate exchangeability holds under the null hypothesis of no treatment effect \citep{sen1966class}.
\item {\it Reflection} \citep{Householder1958}. The Householder matrix $P:= \mathbf{I}_p - 2\mathbf{u}\mathbf{u}^\top$ describes a reflection about the hyperplane orthogonal to the unit vector $\mathbf{u}$. The corresponding group is $\G = \{P, \mathbf{I}_p\}$.
A closely related notion of symmetry is {\it axial} symmetry \citep{Hudecovs2021axial}, where $\G = \{-P, \mathbf{I}_p\}$. 
\item {\it Zonal} symmetry \citep[Example 4]{beran1997multivariate}. This notion of symmetry arises in certain geophysical problems. Let $S^3$ denote the unit sphere in $\R^3$. Then $\G$ consists of all rotations in $\R^3$ that leave the north pole-south pole axis of $S^3$ fixed. Thus, each element of $\G$ corresponds to a rotation in $\R^2$.
\end{enumerate}

In this subsection, we present novel results that combine kernel MMD and OT-based signed-ranks to form new tests for $\G$-symmetry
\begin{equation}\label{eq:G_versus_notG}
    {\rm H}_{\G,0}:{\bf X}  \stackrel{d}{=}  Q  {\bf X}\;\; \forall\;   Q\in \G\ \quad  \text{vs.}\quad \ {\rm H}_{\G,1}:\text{not } {\rm H}_{\G,0}
\end{equation}
that are (i) distribution-free, (ii) universally consistent, and (iii) adapt to a broader class of $\G$-symmetries.

The proposed procedure is motivated by a key observation on the signed-ranks defined via OT: $\X$ is $\G$-symmetric if and only if the population signed-rank of $\X$ follows the known distribution $\nu_S$ (the symmetrized reference distribution, defined in Theorem~\ref{thm:population_rank}).
\begin{theorem}\label{thm:equiv_cond}
Suppose Assumption~\ref{assump:nu-G} holds, $\mathbf{X}\sim \mu\in \mathcal{P}_{\rm ac}(\R^p)$ and $\nu_S$ has a Lebesgue density.
Then the following statements are equivalent:
\begin{enumerate}
\item $\mathbf{X}$ is $\G$-symmetric.
\item $\nabla\psi (\mathbf{X})$ ($\nabla\psi (\cdot )$ is defined in Theorem~\ref{thm:population_rank}) is $\G$-symmetric.
\item $\nabla\psi (\mathbf{X})\sim \nu_S$.
\end{enumerate}
\end{theorem}
A proof of Theorem~\ref{thm:equiv_cond} is given in Appendix~\ref{sec:proof_thm_equiv_cond}. 
With the above fact, we can use the difference (as measured by MMD \citep{Gretton12}) between the empirical distribution of the sample signed-ranks and this known distribution $\nu_S$ as the test statistic to test \eqref{eq:G_versus_notG}. 

Recall the definition of MMD in \eqref{eq:MMD}. If the kernel $K$ is characteristic, MMD is a distance on the space of probability measures \citep{Gretton12}. In particular, ${\rm MMD}(P,Q)\geq 0$, and ${\rm MMD}(P,Q)=0$ if and only if $P=Q$.

From Theorem~\ref{thm:equiv_cond}, $\mathbf{X}$ is $\G$-symmetric if and only if ${\rm MMD}(\nabla\psi(\mathbf{X}),\nu_S)=0$.
The empirical distribution of sample signed-ranks --- denoted by $\frac{1}{n}\sum_{i=1}^n \delta_{\hat{U}_i}$ where $\hat{U}_i :=S_n(\mathbf{X}_i)R_n(\mathbf{X}_i)$ --- provides an approximation for the distribution of the population signed-rank $\nabla\psi (\mathbf{X})$.
Therefore, we propose to use
{\small
\begin{equation*}\label{eq:defnTn}
\begin{aligned}&\quad T_n:={\rm MMD}^2\left(\frac{1}{n}\sum_{i=1}^n \delta_{\hat{U}_i} , \nu_S \right) \\
&=  \frac{1}{n^2}\sum_{i,j=1}^n K(\hat{U}_i,\hat{U}_j) + \E\left[ K(U,U')\right] -\frac{2}{n}\sum_{i=1}^n \E_U\left[K(\hat{U}_i,U) \right],
\end{aligned}
\end{equation*}
}

\noindent as our test statistic for the hypothesis test \eqref{eq:G_versus_notG}. Here $U$, $U'$ are i.i.d.\ from $\nu_S$.

While there could be various choices of the characteristic kernel $K$, here we recommend the use of the Gaussian kernel 
$K(\mathbf{x},\mathbf{y}):=\exp(-\sigma\|\mathbf{x}-\mathbf{y}\|^2)$, as in such a case, $\E_U\left[K(\hat{U}_i,U) \right]$ and $\E\left[ K(U,U')\right]$ have closed form expressions:
\begin{equation*}\label{eq:simple_form_with_Gaussian}\begin{aligned}
\E_U\left[K(\hat{U}_i,U) \right] &= \frac{1}{(2\sigma+1)^{p/2}} \exp \left( - \frac{\sigma}{2\sigma+1}\hat{U}_i^\top \hat{U}_i\right),\\
\E\left[ K(U,U')\right] &= \frac{1}{(4\sigma+1)^{p/2}}.
\end{aligned}\end{equation*}

\begin{remark}[Distribution-Freeness]\label{rk:dist_free}
The test statistic $T_n$ is a function of the OT-based signed-ranks, and consequently has a known distribution under ${\rm H}_{\G,0}$ (see \eqref{eq:G_versus_notG}). Observe that the signed-ranks $(\hat{U}_1,\ldots, \hat{U}_n)$ have the same distribution as $(S_1 R_n(\X_1),\ldots,S_n R_n(\X_n))$, where the absolute ranks $(R_n(\X_1),\ldots,R_n(\X_n))$ are uniformly distributed over the $n!$ permutations of $(\mathbf{h}_1,\ldots, \mathbf{h}_n)$, and is independent of the signs $(S_1,\ldots,S_n)$, with $S_1,\ldots,S_n$ i.i.d.\ uniform over $\G$. Note that this null distribution is distribution-free, i.e., it does not depend on the distribution of $\X$, and consequently, distribution-free critical values can be obtained for testing the null hypothesis~\eqref{eq:G_versus_notG}.
\end{remark}

The following theorem (proved in Appendix~\ref{sec:proof_thm_consistency}) 
establishes the consistency of our test statistic.
\begin{theorem}[Consistency]\label{thm:consistency} Suppose Assumption~\ref{assump:nu-G} holds, and $\mathbf{X}\sim \mu\in \mathcal{P}_{\rm ac}(\R^p)$. Let $\nabla\psi(\cdot)$ be defined as in Theorem~\ref{thm:population_rank}. If the kernel $K(\cdot,\cdot)$ is bounded and continuous, then, as $n \to \infty$,
$$\begin{aligned}
T_n\overset{a.s.}{\longrightarrow} {\rm MMD}^2(\nabla\psi(\mathbf{X}),\nu_S).
\end{aligned}$$
\end{theorem}
From the consistency theorem above, $T_n = o_p(1)$ under ${\rm H}_{\G,0}$, and $T_n = {\rm MMD}(\nabla\psi(\mathbf{X}),\nu_S) + o_p(1)$ under ${\rm H}_{\G,1}$. If the kernel $K(\cdot,\cdot)$ is characteristic, ${\rm MMD}(\nabla\psi(\mathbf{X}),\nu_S) > 0$ under ${\rm H}_{\G,1}$. Therefore, the test that rejects ${\rm H}_{\G,0}$ for large values of $T_n$ is consistent against all fixed alternatives where $\mathbf{X}$ is not $\G$-symmetric.

\begin{remark}[Broader class of $\G$]\label{rk:broader_class}
Theorem~\ref{thm:consistency} does not require $-\mathbf{I}_p\in \G$  which was assumed in the consistency theorem of multivariate Wilcoxon's signed-rank test \cite[Corollary 3.1]{huang2023multivariate}.
Therefore, beyond central, sign, and spherical symmetries, it is also applicable to other notions of symmetry such as exchangeability, reflection, and zonal symmetry.
\end{remark}

To the best of our knowledge, our proposed test for symmetry is the first test that simultaneously satisfies the three aforementioned desirable properties (i.e., distribution-freeness, universal consistency, and applicability to general symmetry).
Numerical experiments for the new method are presented in Appendix~\ref{sec:OT-MMD-simu}, 
which shows that the new test is among the best-performing methods against various types of alternatives.


Observe that our test statistic $T_n$ is not a standard $V$-statistic because the sample signed-ranks are neither mutually independent nor marginally distributed as the population signed-rank. Thus, unlike classical results in the MMD literature, $T_n$ may not exhibit an asymptotic distribution as an infinite mixture of chi-squares. To better characterize its asymptotic behavior, we demonstrate that, after appropriate normalization, $T_n$ satisfies a central limit theorem in the reproducing kernel Hilbert space. 

\subsection{Limiting Behavior of $T_n$ under $\G$-symmetry}\label{sec:recentered_CLT}
In this subsection, we provide a characterization of the asymptotic behavior of the proposed test statistic $T_n$ under the null hypothesis of $\G$-symmetry. In particular, we show that the test statistic $T_n$, after appropriate centering, satisfies a central limit theorem.


Assume that $\mathbf{X}$ is $\G$-symmetric, from which it follows that ${\rm MMD}^2(\nabla\psi(\mathbf{X}),\nu_S)=0$. A natural question to ask is: ``Will $n T_n$ converge in distribution to an infinite mixture of chi-squares (which is the limiting null distribution of the classical MMD for i.i.d.\ data \citep{Gretton12})?''

The answer, however, is negative. Note that $T_n$ is not a standard $V$-statistic like the empirical MMD \citep{gretton2009fast}. Firstly, the sample signed-ranks $\hat{U}_1,\ldots,\hat{U}_n$, which are the products of the signs and absolute ranks, are not independent (as the absolute ranks are a permuted version of $\mathbf{h}_1,\ldots,\mathbf{h}_n$; see Remark~\ref{rk:dist_free}). Secondly, the marginal distribution of $\hat{U}_i$ is not $\nu_S$. In fact, under the null hypothesis~\eqref{eq:G_versus_notG}, $\hat{U}_i$ marginally follows $ \nu_{n,S}$, where $\nu_{n,S}$ denotes the distribution of $S\mathbf{H}_n$ with $S\sim {\rm Uniform}(\mathcal{G})$ independent of $\mathbf{H}_n\sim \nu_n$. The limiting behavior of $nT_n={\rm MMD}^2\big(\frac{1}{n}\sum_{i=1}^n \delta_{\hat{U}_i} , \nu_S \big)$ may depend on how $\nu_{n,S}$ converges to $\nu_S$. Even if $\nu_{n,S}$ approaches $\nu_S$ at $n^{-1/2}$-rate (e.g., $\mathbf{h}_1,\mathbf{h}_2,\ldots$ are drawn i.i.d.\  from $\nu$), $n T_n$ may still diverge (see Remark~\ref{rk:implication_Tn} below). It turns out that in order to have a limiting distribution being an infinite mixture of chi-squares, the suitable centering required for $\frac{1}{n}\sum_{i=1}^n \delta_{\hat{U}_i}$ is $\nu_{n,S}$ instead of $\nu_S$. When centered around $\nu_{n,S}$, $\frac{1}{n}\sum_{i=1}^n \delta_{\hat{U}_i}$ 
follows a central limit theorem, as shown below.

\begin{theorem}\label{thm:recentered_CLT}
Assume that the kernel $K(\cdot,\cdot)$ is bounded and continuous, and $\nu_n $ converges weakly to $\nu$. Let $\mathcal{H}$ denote the RKHS associated with the kernel $K(\cdot,\cdot)$. Under the null hypothesis of $\G$-symmetry, we have
{\small
$$\begin{aligned}
&\sqrt{n}\Big(\frac{1}{n}\sum_{i=1}^n K(\hat{U}_i,\cdot) - \mathbb{E}_{\nu_{n,S}}\left[K(U,\cdot)  \right] \Big)\overset{d}{\longrightarrow} \sum_{i=1}^\infty \sqrt{\lambda_i} \langle \phi_i,\cdot\rangle_\mathcal{H} Z_i,
\end{aligned}
$$
}
and consequently,
{\small
$$\begin{aligned}
n{\rm MMD}^2\Big(\frac{1}{n}\sum_{i=1}^n \delta_{\hat{U}_i} , \nu_{n,S} \Big) &\overset{d}{\longrightarrow} 
\sum_{i=1}^\infty {\lambda_i}Z_i^2,\quad {\rm as\ }n\to\infty.
\end{aligned}
$$
}
Here $Z_i$'s are i.i.d.\ standard normal variables, and $\lambda_i\geq 0$, orthonormal $\{\phi_i\}_{i\geq 1}\subset\mathcal{H}$ ($i\geq 1$) only depend on the kernel $K(\cdot,\cdot)$, $\G$, and $\nu$. 
\end{theorem}

See Appendix~\ref{sec:proof_recentered_CLT} for the proof of Theorem~\ref{thm:recentered_CLT}. In the above result, the $\lambda_i$'s and the $\phi_i$'s are the eigenvalues and eigenfunctions of a trace-class operator over $\mathcal{H}$; see \eqref{eq:Tsigma} and \eqref{eq:sigma_expression} in Appendix~\ref{sec:proof_recentered_CLT} for the exact expression of the operator.
 The proof leverages recent theoretical tools from \cite{fernandez2022general}, which work directly with random functionals on the reproducing kernel Hilbert space to analyze kernel-based tests.

\begin{remark}[Implication for the null distribution of $T_n$]\label{rk:implication_Tn}
Observe that
{\small
$$
\begin{aligned}
n{T}_n&=n{\rm MMD}^2\Big(\frac{1}{n}\sum_{i=1}^n \delta_{\hat{U}_i} , \nu_{S} \Big) \\ 
& =n\Bigg\| \frac{1}{n}\sum_{i=1}^n K(\hat{U}_i,\cdot)  - \mathbb{E}_{U\sim \nu_{S}}K(U,\cdot)\Bigg\|_\mathcal{H}^2\\
&= \Bigg\| \underbrace{\sqrt{n}\Big(\frac{1}{n}\sum_{i=1}^n K(\hat{U}_i,\cdot) - \mathbb{E}_{U\sim \nu_{n,S}}K(U,\cdot)\Big)}_{{\rm converge\ in\ distribution\ by\ Theorem}~\ref{thm:recentered_CLT}}\\
&\quad + \underbrace{\vphantom{\sqrt{n}\left(\frac{1}{n}\sum_{i=1}^n K(\hat{U}_i,\cdot) - \mathbb{E}_{U\sim \nu_{n,S}}K(U,\cdot)\right)}\sqrt{n}\left(\mathbb{E}_{U\sim \nu_{n,S}}K(U,\cdot) - \mathbb{E}_{U\sim \nu_{S}}K(U,\cdot)
\right)}_{\text{\rm non-random}}\Bigg\|_\mathcal{H}^2,
\end{aligned}
$$
}
where the first term above converges in distribution by Theorem~\ref{thm:recentered_CLT}, and the second term is non-random. If $\sqrt{n}\left(\mathbb{E}_{U\sim \nu_{n,S}}K(U,\cdot) - \mathbb{E}_{U\sim \nu_{S}}K(U,\cdot)
\right)$ converges to $0\in\mathcal{H}$ as $n\to\infty$, then $nT_n$ will have the same limiting distribution $\sum_{i=1}^\infty {\lambda_i}Z_i^2$ as in Theorem~\ref{thm:recentered_CLT}. However, this requirement on the reference distribution may be too strong and hard to obtain. For example, if $\mathbf{h}_1,\mathbf{h}_2,\ldots$ are i.i.d.\ from $\nu$ (and fixed), then by the law of the iterated logarithm \cite[Theorem~4.1]{Kuelbs1977lil}, $\sqrt{n}\big(\mathbb{E}_{U\sim \nu_{n,S}}K(U,\cdot) - \mathbb{E}_{U\sim \nu_{S}}K(U,\cdot)
\big)$ will be of order $\sqrt{\log\log n}$, and will not converge to 0 or any constant vector.
\end{remark}
It may seem appealing from Theorem~\ref{thm:recentered_CLT} to use the statistic $n{\rm MMD}^2\big(\frac{1}{n}\sum_{i=1}^n \delta_{\hat{U}_i} , \nu_{n,S} \big)$ for testing the hypothesis of $\G$-symmetry, 
as this statistic is not only distribution-free in finite samples, but also has a known asymptotic distribution (the infinite mixture of chi-squares).
However, computing ${\rm MMD}^2\big(\frac{1}{n}\sum_{i=1}^n \delta_{\hat{U}_i} , \nu_{n,S} \big)$ is typically more difficult than $T_n={\rm MMD}^2\big(\frac{1}{n}\sum_{i=1}^n \delta_{\hat{U}_i} , \nu_S \big)$, as $\E_{U,U'\sim \nu_{n,S}}\big[ K(U,U')\big]$ and $ \E_{U\sim\nu_{n,S}} \big[K(\hat{U}_i,U) \big]$ may not have a simple closed form. The computational complexity to evaluate them can be high, especially when $|\G|$ is large or infinite. Thus, from a practical perspective, we recommend using the statistic $T_n$.

\section{Recent Literature Review}\label{sec:review-further}
In this section, we briefly sketch some further applications of OT in nonparametric testing.
It will be seen that many of these methods create distribution-free tests by substituting the data points in some classical test statistics with their corresponding OT-based ranks. 
Note that when an asymptotic level $\alpha$ test $\phi^{(n)}$ is based on the distribution-free OT-based ranks and signs, i.e., $\lim_{n\to\infty}\E_\mathbb{P} [\phi^{(n)}]\leq \alpha$, then we automatically have $\lim_{n\to\infty}\sup_{\mathbb{P}\in \mathcal{P}}\E_\mathbb{P} [\phi^{(n)}]\leq \alpha$, where $\mathcal{P}$ is the class of data distributions for which the distribution-free property holds true.
Such a uniform inferential validity property is quite desirable. 

For the {\it two-sample problem} that has been introduced in Section~\ref{sec:GWRS},
instead of considering the statistic $T^{\nu}_{m,n}$,
~\cite{boeckel2018multivariate} used the 2-Wasserstein distance between the empirical distributions $\frac{1}{m}\sum_{i=1}^m \delta_{\hat{R}_{m,n}(\X_i)}$ and $\frac{1}{n}\sum_{i=1}^n \delta_{\hat{R}_{m,n}(\Y_i)}$ as the test statistic; \cite{Nabarun2021rank} used the {\it energy distance} \citep{szekely2013energy} betweeen the empirical distributions $\frac{1}{m}\sum_{i=1}^m \delta_{\hat{R}_{m,n}(\X_i)}$ and $\frac{1}{n}\sum_{i=1}^n \delta_{\hat{R}_{m,n}(\Y_i)}$ as the test statistic; and \cite{deb2021efficiency} used kernel MMD distance \citep{Gretton12} as described in Section~\ref{sec:RankMMD}.

A special choice of the reference distribution $\nu$ and $\nu_n$ (see Section~\ref{sec:OT-multi-ranks}) was proposed in \cite{hallin2017distribution} and adopted in some following works \citep{shi2020rate,Shi2022indep,hallin2023fully,shi2023semiparametrically}.
In \cite{hallin2017distribution}, $\nu$ is set as the spherical uniform distribution, and $\nu_n$ is carefully selected as a uniform distribution over a regular grid. The construction of the grid (over the unit ball) starts by factorizing $n$ into $n=n_R n_S + n_0$ with $n_R,n_S\to\infty$ and $n_0/n\to 0$ as $n\to\infty$. $n_S$ distinct unit vectors $\mathbf{s}_1,\ldots,\mathbf{s}_{n_S}$ are chosen on the unit hypersphere to be as regular as possible. 
The grid then contains the $n_R n_S$ points $\frac{r}{n_R+1} \mathbf{s}_s$, $r=1,\ldots,n_R$, $s=1,\ldots,n_S$.
The remaining $n_0$ points are either $n_0$ copies of the origin \citep{hallin2017distribution}, or a random sample from $\frac{1}{2(n_R+1)} \mathbf{s}_1,\ldots, \frac{1}{2(n_R+1)} \mathbf{s}_{n_S}$ without replacement \citep{shi2020rate}.
$(n_R+1)\|\hat{R}_n(\X_i)\|$ and $\hat{R}_n(\X_i)/ \|\hat{R}_n(\X_i)\|$ are referred to as the {\it center-outward rank} and the {\it center-outward sign} of $\X_i$, respectively.

\cite{hallin2023fully} applied OT-based ranks in the multiple-output linear regression model, where each observation follows the model
$$\Y_i = {\B \beta}_0 + {\B \beta}^\top \mathbf{c}_i + {\B \varepsilon}_i, \qquad \mbox{for }\; i = 1,\ldots, n.$$
Here, $\Y_i \in \R^p$, ${\B \beta}_0\in\R^p$ is the intercept, ${\B \varepsilon}_i \in \R^p$ is the unobserved error, $\mathbf{c}_i\in\R^m $ are known covariates, and ${\B \beta}\in \R^{p\times m}$ are the regression coefficients.
When $\mathbf{c}_i$ is the one-hot encoding of a class label (i.e., a $(K-1)$-dimensional vector with at most one entry equal to 1 and all others equal to 0),
testing whether ${\B \beta} =\mathbf{0}_{(K-1)\times p}$ corresponds to the classical $K$-sample location problem or the MANOVA model.
The test statistic proposed by \cite{hallin2023fully} is based on the OT ranks of $(\mathbf{Z}_1,\ldots,\mathbf{Z}_n)$, where $\mathbf{Z}_i := \Y_i - {\B \beta}^\top \mathbf{c}_i$. Let $\bar{\mathbf{c}}$ be $ n^{-1}\sum_{i=1}^n \mathbf{c}_i$, $(\hat{R}(\mathbf{Z}_1),\ldots, \hat{R}(\mathbf{Z}_n))$ be the OT ranks of $(\mathbf{Z}_1,\ldots,\mathbf{Z}_n)$, and $J(\cdot)$ be some score function. The test statistic for the hypothesis that ${\B \beta}$ equals a specified value is given by
$$\Lambda_J := n^{-1} \sum_{i=1}^n \mathbf{K}_{\mathbf{c}}^{(n)\top} (\mathbf{c}_i - \bar{\mathbf{c}})J^\top (\hat{R}(\mathbf{Z}_i)),$$
where $\mathbf{K}_{\mathbf{c}}^{(n)}$ is an $m\times m$ matrix that depends on $\{\mathbf{c}_i\}_{i=1}^n$ (and its limiting behavior; see \cite[Section 4.2]{hallin2023fully}).
$\Lambda_J$ is asymptotically normal, and the test based on $\Lambda_J$ has a locally asymptotically maximin property if the score function $J(\cdot)$ is suitably chosen \citep[Proposition~5.1]{hallin2023fully}.
 In the case of two-sample location problem, $\Lambda_J$ with $J(\x)=\x$ reduces to the Wilcoxon's rank-sum test statistics discussed in Section~\ref{sec:GWRS}.

Let us now consider the problem of {\it independence testing}. Suppose we would like to test whether two random vectors $\X$ and $\Y$ are independent, given i.i.d.\ observations $(\X_1,\Y_1),\ldots$, $(\X_n,\Y_n)$ from the same distribution.
A practical approach to constructing distribution-free tests is to apply a classical test statistic to the OT-based rank transformed data.
\cite{Nabarun2021rank,Shi2022indep} first rank transform $\{\X_i\}_{i=1}^n$ and $\{\Y_i\}_{i=1}^n$ to obtain their OT-based ranks $\{\hat{R}_n^{\X}(\X_i)\}_{i=1}^n$ and $\{\hat{R}_n^{\Y}(\Y_i)\}_{i=1}^n$.
The {\it distance covariance} \citep{szekely2007distance,szekely2013distance} between $\{\hat{R}_n^{\X}(\X_i)\}_{i=1}^n$ and $\{\hat{R}_n^{\Y}(\Y_i)\}_{i=1}^n$ is then used as the test statistic, which leads to a distribution-free test that is consistent against all fixed alternatives.

\cite{shi2020rate} introduced a class of {\it generalized symmetric covariances} (GSCs) for independence testing, which includes distance covariance, multivariate versions of Hoeffding's $D$, Blum-Kiefer-Rosenblatt's $R$ as special cases. Along with GSCs, they proposed a OT rank-based version which is distribution-free with certain consistency and transformation invariance properties. Further, they analyzed the local power of their proposed tests, under the setting of a parametrized family called {\it Konijn} alternatives where
$$\X = \begin{pmatrix} \X_1 \\ \X_2 \end{pmatrix} := \begin{pmatrix} \mathbf{I}_{d_1} & \delta \mathbf{M}_1 \\ \delta \mathbf{M}_2 & \mathbf{I}_{d_2}\\ \end{pmatrix} \begin{pmatrix} \X_1^* \\ \X_2^* \end{pmatrix} = \mathbf{A}_\delta \begin{pmatrix} \X_1^* \\ \X_2^* \end{pmatrix}$$
for some unobserved independent random vectors $ \X_1^*$ and $\X_2^*$. The proposed test was shown to to achieve rate optimality (with root-$n$ rate).

\cite{shi2023semiparametrically} proposed center-outward sign, Spearman, Kendall, and score tests based on $\{\hat{R}_n^{\X}(\X_i)\}_{i=1}^n$ and $\{\hat{R}_n^{\Y}(\Y_i)\}_{i=1}^n$ as well. These tests reject the hypothesis of independence for large Frobenius norms of
$$\begin{aligned}
\mathbf{W}_{\rm sign}^{(n)} &:= \frac{1}{n}\sum_{i=1}^n \frac{\hat{R}_n^{\X}(\X_i)}{\|\hat{R}_n^{\X}(\X_i)\|} \frac{\hat{R}_n^{\Y}(\Y_i)^\top}{ \|\hat{R}_n^{\Y}(\Y_i)\|},\\
\mathbf{W}_{\rm Spearman}^{(n)} &:= \frac{1}{n}\sum_{i=1}^n \hat{R}_n^{\X}(\X_i) \hat{R}_n^{\Y}(\Y_i)^\top, \\
\mathbf{W}_{\rm Kendall}^{(n)} &:= \begin{pmatrix} n\\ 2 \end{pmatrix}^{-1}\sum_{i<j} {\rm sign} \big[ \delta_{i,j}^\X (\delta_{i,j}^\Y)^\top \big],
\end{aligned}$$
with $\delta_{i,j}^\X :=  \hat{R}_n^{\X}(\X_i) - \hat{R}_n^{\X}(\X_j)$, $\delta_{i,j}^\Y = \hat{R}_n^{\Y}(\Y_i) -  \hat{R}_n^{\Y}(\Y_j)$, and
$$\begin{aligned}
\mathbf{W}_{\mathbf{J}}^{(n)} &:= \frac{1}{n}\sum_{i=1}^n J_1(\|\hat{R}_n^{\X}(\X_i)\|) J_2(\|\hat{R}_n^{\Y}(\Y_i)\|)\\
&\qquad \qquad \cdot \frac{\hat{R}_n^{\X}(\X_i)}{\|\hat{R}_n^{\X}(\X_i)\|} \frac{\hat{R}_n^{\Y}(\Y_i)^\top }{\|\hat{R}_n^{\Y}(\Y_i)\|}
\end{aligned}$$
respectively, where the sign$(\cdot)$ in the definition of $\mathbf{W}_{\rm Kendall}^{(n)}$ is the element-wise sign that applies to each entry of the matrix. 
The ARE of the tests based on $\mathbf{W}_{\mathbf{J}}^{(n)}$ and $\mathbf{W}_{\rm Spearman}^{(n)}$ relative to Wilks' test is analyzed within the framework of the generalized Konijn family \cite[Section 4.1]{shi2023semiparametrically}. This analysis also establishes a Chernoff-Savage-type property.

\cite{del2024nonparametric} applied OT in nonparametric multiple-output quantile regression. They defined the {\it conditional center-outward quantile map} as the OT map from a spherical uniform reference distribution to the conditional distribution of $\Y$ given $\X=\x$.
Let $\bar{\mathbb{S}}_p$ denote the closed unit ball and $\mathcal{S}_{p-1}$ denote the unit hypersphere in $\R^p$.
The {\it conditional center-outward quantile regions} and {\it contours} were defined as the image of $\tau \bar{\mathbb{S}}_p $ and $\tau \mathcal{S}_{p-1}$, respectively, under the conditional quantile map, where $\tau \in (0,1)$ represents a quantile level.
They also proposed consistent estimation methods for these conditional quantile concepts based on i.i.d.\ observations.


OT-based ranks have also been employed for R-estimation to estimate the parameters of semiparametric {\it vector autoregressive moving average} (VARMA) time series models \citep{Hallin2022VARMA}. Additionally, \cite{Hallin2023VAR} developed robust testing procedures using OT-based ranks for semiparametric {\it vector autoregressive} (VAR) models, addressing inferential challenges such as selecting the autoregressive order in a VAR model and testing for serial dependence in multiple-output regression.


\cite{Hallin2024directional} addressed the analysis of directional data, i.e., data taking values on the unit hypersphere $\mathcal{S}^{p-1}$. Their proposed method computes the OT-based rank map by solving an OT problem from the data distribution to a reference distribution, termed the {\it directional distribution function} in their work. The reference distribution is chosen as the uniform distribution over $\mathcal{S}^{p-1}$. Unlike the standard OT approach using squared Euclidean distance (as in \eqref{eq:OT_highdim}), they employ the squared geodesic distance, defined as 
$c(\x,\x')=|{\rm arccos}(\x^\top \x')|^2/2$, for $\x,\x' \in \R^p$. Based on the directional distribution function, they introduce directional quantile functions, quantile contours, quantile regions, and data-driven systems of (curvilinear) parallels and (hyper)meridians. Leveraging this framework, they develop a universally consistent test for uniformity, which assesses whether the data distribution is uniform over $\mathcal{S}^{d-1}$. Additionally, they propose a class of distribution-free and consistent tests for directional MANOVA, designed to test whether $m\geq 2$ independent samples on the unit hypersphere originate from the same distribution.


\cite{ghosal2022multivariate} took a slightly different approach to defining multivariate ranks and quantiles via OT, avoiding the need to discretize the reference distribution. Rather than solving a discrete-discrete OT problem that maps $n$ data points to $n$ reference rank vectors, they formulated a {\it semi-discrete OT} problem, transporting $n$ data points to a known continuous reference distribution. This method results in non-unique ranks, where the rank of a data point is not a single point but a region in $\R^p$ with mass $1/n$. They established the uniform convergence of the empirical quantile and rank maps and analyzed their rate of convergence. Furthermore, this concept of OT-based ranks can also be used in two-sample testing and independence testing~\cite[Section 6]{ghosal2022multivariate}.

OT based ideas (in particular, the 2-Wasserstein distance) was also applied in \cite{Mordant2022dep} to define dependence coefficients that take values between 0 and 1; also see~\cite{deb2024}. 

\section{Open Problems}\label{sec:Open}


In this final section, we highlight a few open questions that arise from the discussions in this paper.

While the multivariate (generalized) Wilcoxon's rank-sum and signed-rank tests extend the Hodges-Lehmann and Chernoff-Savage lower bounds to the multivariate setting, the current results remain restricted to specific subclasses of distributions, such as those with independent components or elliptically symmetric distributions. Whether these results hold in more generality remains an open question. From a technical perspective, the main challenge in extending these results lies in the lack of a precise characterization of the OT map from $\mu$ to $\nu$ in more general settings.


OT-based signs and signed-ranks possess many desirable properties, such as distribution-freeness under the hypothesis of $\G$-symmetry, making them useful for constructing powerful tests for $\G$-symmetry (or its location-shift subproblem), as discussed in Section~\ref{sec:multi-signed-rank-OT}. However, other widely studied notions of symmetry, such as elliptical symmetry\footnote{An elliptically symmetric random vector is an affine transformation (i.e., $\x\mapsto A\x +\mathbf{b}$) of a spherically symmetric random vector.}, do not naturally fit within the $\G$-symmetry framework. In fact, under the assumption of elliptical symmetry, it has been shown that Mahalanobis ranks and signs are important invariant statistics which can be used to design optimal tests for location and shape parameters \citep{Hallin2002optimal,Hallin2006shape}. Whether the theory of OT can provide distribution-free methods for testing elliptical symmetry or lead to powerful tests in this setting remains an open problem.


The two-sample test based on OT ranks and kernel MMD (see Section~\ref{sec:RankMMD}) has non-trivial Pitman efficiency against $O(1/\sqrt{m+n})$ alternatives \citep[Theorem 4.3]{deb2021efficiency}. A natural question is whether the one-sample test based on OT signed-ranks and kernel MMD (see Sections~\ref{sec:one-sample-kernel-OT} and~\ref{sec:recentered_CLT}) also achieves non-trivial Pitman efficiency. In the two-sample problem, it was shown that the OT-based sample ranks $\hat{R}_n(\X_i)$, appearing in the test statistic, can be replaced by their oracle population counterparts $R(\X_i)$ without affecting the asymptotic null distribution (a H{\' a}jek-type representation result),  using the permutation distribution under the null \citep[Theorem C.1]{deb2021efficiency}. However, this permutation technique is not applicable in the one-sample setting. It remains unknown whether a similar H{\' a}jek representation result holds in the one-sample problem and, consequently, whether nontrivial Pitman efficiency can be established in this context.

\begin{funding}
B.\ Sen is supported by NSF Grant DMS-2311062.
%
\end{funding}

\begin{appendix}

\section{Proofs}\label{sec:AppendixProofs}
\subsection{Proof of Theorem~\ref{thm:equiv_cond}}\label{sec:proof_thm_equiv_cond}
1.\ $\Rightarrow$ 2. Fix $S \in \G$.
If $\mu$ is $\G$-symmetric, then $\nabla\psi (\mathbf{X})\overset{d}{=}\nabla\psi (S\mathbf{X})=S\nabla\psi (\mathbf{X})$, where the last equality follows from the equivariance of $\nabla\psi(\cdot)$ \citep[Theorem 2.1-(iii)]{huang2023multivariate}. Thus, $\nabla\psi (\mathbf{X})$ is $\G$-symmetric.

2.\ $\Rightarrow$ 1. If $\nabla\psi \# \mu$ is $\G$-symmetric, then for all $Q\in\G$, 
\begin{equation}\label{eq:nabla_eqdist}\nabla\psi (\mathbf{X})\overset{d}{=} Q\nabla\psi (\mathbf{X})
=\nabla\psi (Q\mathbf{X}).
\end{equation}
From Theorem~\ref{thm:population_rank}-(ii), 
$\nabla \psi(\cdot)$ is the $\mu_S$-a.e.\ unique gradient of convex function that pushes $\mu_S$ to $\nu_S$.
Since $\nu_S$ has a Lebesgue density (from theorem assumption), McCann's theorem \citep[Remark 16]{McCann1995} states that $\nabla \psi(\cdot )$ admits an almost sure inverse $\nabla\psi^*:\R^p\to\R^p$ such that
$\nabla\psi^*(\nabla\psi(\mathbf{x}))=\mathbf{x}$ (for $\mu_S$ a.e.-$\mathbf{x}$) and $\nabla\psi \circ \nabla\psi^* (\mathbf{x}) = \mathbf{x}$ (for $\nu_S$ a.e.-$\mathbf{x}$).
From \cite[Lemma B.3]{huang2023multivariate}, $\mu$ is absolutely continuous w.r.t.\ $\mu_S$.
Therefore, $\nabla\psi^*(\nabla\psi(\mathbf{x}))=\mathbf{x}$ for $\mu$ a.e.-$\mathbf{x}$ as well. Similarly, 
as the distribution of $Q\mathbf{X}$ is absolutely continuous w.r.t.\ $\mu_S$,
we also have $\nabla\psi^*(\nabla\psi(Q\mathbf{x}))=Q\mathbf{x}$ for $\mu$ a.e.-$\mathbf{x}$.
Applying $\nabla\psi^*$ on both sides of \eqref{eq:nabla_eqdist}, we have $\mathbf{X}\overset{d}{=}Q\mathbf{X}$.

2.\ $\Rightarrow$ 3. If $\nabla\psi (\mathbf{X})$ is $\G$-symmetric, then with $S\sim {\rm Uniform}(\G)$ independent of $\mathbf{X}$, we have $\nabla\psi (\mathbf{X})\overset{d}{=} S\nabla\psi (\mathbf{X}) = \nabla\psi (S\mathbf{X})\sim \nabla\psi \# \mu_S = \nu_S$; the last equality follows from \cite[Theorem 2.1-(ii)]{huang2023multivariate}.

3.\ $\Rightarrow$ 2. Since $\nu_S$ is $\G$-symmetric, if $\nabla\psi (\mathbf{X})\sim \nu_S$, then $\nabla\psi (\mathbf{X})$ is $\G$-symmetric.

\qed

\subsection{Proof of Theorem~\ref{thm:consistency}}\label{sec:proof_thm_consistency}
Fix $\omega\in \Omega$.
Let $\mathcal{X}_n$ follow the empirical distribution over $\mathbf{X}_1,\ldots,\mathbf{X}_n$.
For almost every $\omega\in \Omega$, we have $(\mathcal{X}_n,S_n(\mathcal{X}_n)R_n(\mathcal{X}_n))\overset{d}{\longrightarrow}(\mathbf{X},\nabla\psi(\mathbf{X}))$, as $n \to \infty$ \citep[Lemma B.5]{huang2023multivariate}.

Let $\mathcal{X}_n'$ be an independent copy of $\mathcal{X}_n$. Then we have
$$\begin{aligned}(\mathcal{X}_n,S_n(\mathcal{X}_n)R_n(\mathcal{X}_n),\mathcal{X}_n',S_n(\mathcal{X}_n')R_n(\mathcal{X}_n'))\\
\overset{d}{\longrightarrow}(\mathbf{X},\nabla\psi(\mathbf{X}), \mathbf{X}',\nabla\psi(\mathbf{X}')).
\end{aligned}$$

Consider the continuous function given by
$$f(\mathbf{x}_1,\mathbf{y}_1,\mathbf{x}_2,\mathbf{y}_2 ) :=\left| K(\mathbf{y}_1,\mathbf{y}_2) - K(\nabla\psi(\mathbf{x}_1),\nabla\psi(\mathbf{x}_2))\right|.$$
If $K(\cdot,\cdot)$ is continuous, then the continuous mapping theorem implies, as $n \to \infty$,
{\small$$\begin{aligned}&\left| K\big(S_n(\mathcal{X}_n)R_n(\mathcal{X}_n),S_n(\mathcal{X}_n')R_n(\mathcal{X}_n')\big) - K(\nabla\psi(\mathcal{X}_n),\nabla\psi(\mathcal{X}_n'))\right|\\
&\quad \overset{d}{\longrightarrow}0.\end{aligned}$$}
The boundedness of $K$ then implies the following expectation also converges:
$$\frac{1}{n^2} \sum_{i,j=1}^n |K(\hat{U}_i,\hat{U}_j) - K(\nabla\psi(\mathbf{X}_i),\nabla\psi(\mathbf{X}_j))| \to 0.$$

Similarly, by considering $(\mathcal{X}_n,S_n(\mathcal{X}_n)R_n(\mathcal{X}_n))\overset{d}{\longrightarrow}(\mathbf{X},\nabla\psi(\mathbf{X}))$ and the continuous function
$$f(\mathbf{x},\mathbf{y}) := \left| \E_U\left[K(\mathbf{y},U) \right]  - \E_U\left[K(\nabla\psi(\mathbf{x}),U) \right] \right|,$$
we have
{\small
$$ \left| \E_U\left[K(S_n(\mathcal{X}_n)R_n(\mathcal{X}_n),U) \right]  - \E_U\left[K(\nabla\psi(\mathcal{X}_n),U) \right] \right|\overset{d}{\longrightarrow}0.$$}
The boundedness of $K$ implies the expectation also converges:
$$\frac{1}{n} \sum_{i=1}^n \left| \E_U\left[K(\hat{U}_i,U) \right] - \E_U\left[K(\nabla\psi(\mathbf{X}_i),U) \right]\right|\to 0.$$
Therefore,  as $n \to \infty$,
{\small
$$\begin{aligned}
T_n&=\frac{1}{n^2}\sum_{i,j=1}^n K(\nabla\psi(\mathbf{X}_i),\nabla\psi(\mathbf{X}_j)) + \E\left[ K(U,U')\right] \\
&\quad -\frac{2}{n}\sum_{i=1}^n \E_U\left[K(\nabla\psi(\mathbf{X}_i),U) \right] + o(1)\\
&\longrightarrow {\rm MMD}(\nabla\psi(\mathbf{X}),\nu_S).
\end{aligned}$$
}
Note that we have fixed $\omega\in \Omega$, and the above argument holds for almost every $\omega\in\Omega$. Therefore,
$T_n$ converges almost surely to ${\rm MMD}(\nabla\psi(\mathbf{X}),\nu_S)$. 
\qed

\subsection{Proof of Theorem~\ref{thm:recentered_CLT}}\label{sec:proof_recentered_CLT}
Let $S_i$ be i.i.d.\ ${\rm Uniform}(\mathcal{G})$. Under the null distribution, by the distribution-free property of OT signed-ranks \citep[Proposition 2.2]{huang2023multivariate},
{\small $$
\begin{aligned}
&\quad n{\rm MMD}^2\left(\frac{1}{n}\sum_{i=1}^n \delta_{\hat{U}_i} , \nu_{n,S} \right)\\
&= n\bigg\| \frac{1}{n}\sum_{i=1}^n K(\hat{U}_i,\cdot) - \mathbb{E}_{U\sim \nu_{n,S}}K(U,\cdot)\bigg\|_\mathcal{H}^2 \\
&\overset{d}{=}  \bigg\| \frac{1}{\sqrt{n}}\sum_{i=1}^n K(S_i \mathbf{h}_i,\cdot) - \mathbb{E}\Big[ \frac{1}{\sqrt{n}}\sum_{i=1}^n K(S_i \mathbf{h}_i,\cdot) \Big]\bigg\|_\mathcal{H}^2.
\end{aligned}
$$}
It suffices to show that
{\scriptsize
$$\frac{1}{\sqrt{n}}\sum_{i=1}^n K(S_i \mathbf{h}_i,\cdot) - \mathbb{E}\Big[ \frac{1}{\sqrt{n}}\sum_{i=1}^n K(S_i \mathbf{h}_i,\cdot) \Big] \overset{d}{\longrightarrow} \sum_{i=1}^\infty \sqrt{\lambda_i}\langle \phi_i,\cdot\rangle_{\mathcal{H}}Z_i.$$}
We will apply the following theorem from \cite[Theorem 1]{fernandez2022general} to establish the asymptotic normality of 
{\small \begin{equation}\label{eq:S_n}
S_n(\cdot):=\frac{1}{\sqrt{n}}\sum_{i=1}^n K(S_i \mathbf{h}_i,\cdot) - \mathbb{E}\Big[ \frac{1}{\sqrt{n}}\sum_{i=1}^n K(S_i \mathbf{h}_i,\cdot) \Big].
\end{equation}}

\begin{theorem}\label{thm:kernel_CLT_tool}
Let $(S_n)_{n\geq 1}$ be a sequence of bounded linear test statistics, where each $S_n : \mathcal{H} \to \mathcal{H}$ is defined from the RKHS $\mathcal{H}$ to itself. Suppose the  the following conditions hold:
\begin{enumerate}
\item[$G_0$:] There exists a continuous bilinear form\footnote{Note that for $\sigma(w,w)$ defined only on the diagonal, it can be extended to a bilinear operator via $\sigma(w,w'):=\frac{1}{2}\big(\sigma(w+w',w+w')-\sigma(w,w) - \sigma(w',w')\big)$.} $\sigma:\mathcal{H}\times\mathcal{H}\to\R$ such that for any $w\in\mathcal{H}$, $S_n(w)\overset{d}{\longrightarrow}\mathcal{N}\left(0,\sigma(w,w)\right)$ as $n\to\infty$. 
\item[$G_1$:] For some orthonormal basis $(\phi_i)_{i\geq 1}$ of $\mathcal{H}$, we have $\sum_{i\geq 1}\sigma(\phi_i,\phi_i)<\infty$.
\item[$G_2$:] For some orthonormal basis $(\phi_i)_{i\geq 1}$ of $\mathcal{H}$, and any $\varepsilon>0$, we have
$$\lim_{i\to\infty}\limsup_{n\to\infty}\mathbb{P}\left( \sum_{k=i}^\infty S_n(\phi_k)^2\geq \varepsilon\right)=0.$$
\end{enumerate}
Let $\sigma$ be the bilinear form in condition $G_0$. Define $T_\sigma:\mathcal{H}\to\mathcal{H}$ as
\begin{equation}\label{eq:Tsigma}
    (T_\sigma w)(x)=\sigma(w,K(x,\cdot)).
\end{equation}
Then $T_\sigma$ is self-adjoint and trace class. Suppose $(\lambda_i,\phi_i)_{i\geq 1}$ are the eigenvalues and eigenvectors of $T_\sigma$. Define $S(\cdot)=\sum_{i=1}^\infty \sqrt{\lambda_i}\langle \phi_i,\cdot\rangle_\mathcal{H} Z_i$, where $(Z_i)_{i\geq 1}$ are i.i.d.\ standard normal random variables. We have
$$ S_n\overset{d}{\longrightarrow}S,\quad {\rm and}\quad  \|S_n\|_{\mathcal{H}\to\R}^2\overset{d}{\longrightarrow}\sum_{i=1}^\infty \lambda_iZ_i^2.$$
\end{theorem}
Note that $S_n$ belongs to $\mathcal{H}^*$, the space of bounded linear functionals on $\mathcal{H}$. By the Riesz representation theorem, $\mathcal{H}$ and $\mathcal{H}^*$ are isometric. Thus, the standard operator norm $\|\cdot\|_{\mathcal{H}\to\R}$ is the same as the RKHS norm $\|\cdot\|_{\mathcal{H}}$. The weak convergence of $S_n$ to $S$ in this context may be thought of as the combination of finite-dimensional convergence (conditions $G_0$, $G_1$) and tightness (condition $G_2$); see \cite[Theorem 4.2]{Billingsley1968}.


We first verify condition $G_0$. We will establish the asymptotic normality of $S_n(w)$ with Lindeberg-Feller's CLT.
For any $w\in\mathcal{H}$, let 
$${x}_{n,i} :=\frac{1}{\sqrt{n}}\left( w(S_i\mathbf{h}_i) - \E w(S_i\mathbf{h}_i)\right).$$
Since the kernel is bounded, every function in the RKHS must be bounded because
$|w(x)| = |\langle w,K(x,\cdot)\rangle_\mathcal{H} |\leq \|w\|_\mathcal{H}\cdot \sqrt{\sup_x K(x,x)}$ for all $x$.
Therefore, there exists a constant $C>0$ such that $|{x}_{n,i}| < C/\sqrt{n}$ for all $n$ and $i$. For every $\varepsilon>0$, when $n > C^2/\varepsilon^2$, we have $|{x}_{n,i}|< \varepsilon$.
This implies
$$\begin{aligned}
\lim_{n\to\infty}\sum_{i=1}^n \E \left[| {x}_{n,i}|^2 1_{|{x}_{n,i}|\geq \varepsilon}\right] = 0.
\end{aligned}$$
Hence, by Lindeberg-Feller CLT,
$$S_n(w) = \sum_{i=1}^n {x}_{n,i}\overset{d}{\longrightarrow} \mathcal{N}({0}, \sigma^2),$$
where
{\small
$$\begin{aligned}
\sigma^2&=\lim_{n\to\infty}\E \left[\frac{1}{n}\sum_{i=1}^n \big[ w(S_i \mathbf{h}_i)-\E w(S_i \mathbf{h}_i) \big]^2   \right],
\end{aligned}$$
}
Let $S,S',S''$ be i.i.d.\ from ${\rm Uniform}(\mathcal{G})$.
Define $$f_w(\x) := \E\big[\big[w(S \x)-\E w(S' \x) \big]\big[ w(S \x)-\E w(S'' \x) \big] \big].$$
Then, the above $\sigma^2$ can be written as
$$\begin{aligned}
\sigma^2&=\lim_{n\to\infty}\frac{1}{n} \sum_{i=1}^n f_w(\mathbf{h}_i).
\end{aligned}$$
Since $\frac{1}{n}\sum_{i=1}^n \delta_{\mathbf{h}_i}\overset{d}{\longrightarrow}\nu$ and
$f_w$ is a bounded continuous function, we have
{\small
$$\begin{aligned}
\sigma^2&=\lim_{n\to\infty}\frac{1}{n} \sum_{i=1}^n f_w(\mathbf{h}_i)=\E f_w(\mathbf{H}),
\end{aligned}$$
}
where $\mathbf{H}\sim \nu$.
Define a bilinear form
\begin{equation}\label{eq:sigma_expression}
    \begin{aligned}
&\ \sigma(w_p,w_q):=\E_{S,\mathbf{H}}\Big[\big[w_p(S \mathbf{H})-\E_{S'} [w_p(S' \mathbf{H})] \big]\\
&\qquad \qquad \qquad\cdot \big[ w_q(S \mathbf{H})-\E_{S''} [w_q(S'' \mathbf{H}) ]\big] \Big].
\end{aligned}
\end{equation}
It can be seen that $\sigma^2=\sigma(w,w)$.
Thus, condition~$G_0$ holds.

Next, we verify condition~$G_1$. Let $(\phi_i)_{i\geq 1}$ be an orthonormal basis of $\mathcal{H}$. We have
$$\begin{aligned}
\sum_{i\geq 1} \sigma(\phi_i,\phi_i) &=\sum_{i\geq 1} \E_{S,\mathbf{H}}\left[ (\phi_i(S\mathbf{H})-\E_{S'}\phi_i(S'\mathbf{H}))^2 \right]\\
&\leq \sum_{i\geq 1} \E_{S,\mathbf{H}}\left[ \phi_i(S\mathbf{H})^2 \right]\\
& = \sum_{i\geq 1} \E_{S,\mathbf{H}}\left[ \langle \phi_i, K(S\mathbf{H},\cdot)\rangle_{\mathcal{H}}^2 \right]\\
& = \E_{S,\mathbf{H}}\left[ \| K(S\mathbf{H},\cdot)\|_{\mathcal{H}}^2 \right] \\
& = \E_{S,\mathbf{H}}\left[ K(S\mathbf{H},S\mathbf{H}) \right]<\infty.
\end{aligned}$$
Hence, condition~$G_1$ holds.

Finally, we verify condition~$G_2$.
By Markov inequality, we have
$$\begin{aligned}
&\ \mathbb{P}\Big( \sum_{k=i}^\infty S_n(\phi_k)^2\geq \varepsilon\Big) \leq 
\varepsilon^{-1}\sum_{k=i}^\infty \E\left[S_n(\phi_k)^2 \right]\\
&=\varepsilon^{-1}\sum_{k=i}^\infty \E\Big( \Big\{\frac{1}{\sqrt{n}}\sum_{j=1}^n [\phi_k(S_j \mathbf{h}_j) - 
\E_{S'} \phi_k(S' \mathbf{h}_j) ]\Big\}^2\Big)\\
&=\varepsilon^{-1}\sum_{k=i}^\infty \E\Big( \frac{1}{n}\sum_{j=1}^n [\phi_k(S_j \mathbf{h}_j) - 
\E_{S'} \phi_k(S' \mathbf{h}_j) ]^2\Big)\\
&\leq \varepsilon^{-1}\sum_{k=i}^\infty \E\Big( \frac{1}{n}\sum_{j=1}^n [\phi_k(S_j \mathbf{h}_j)]^2\Big)\\
&=\varepsilon^{-1}\sum_{k=i}^\infty  \frac{1}{n}\sum_{j=1}^n \E_S\left( [\phi_k(S \mathbf{h}_j)]^2\right)\\
&=\varepsilon^{-1}\sum_{k=i}^\infty  \frac{1}{n}\sum_{j=1}^n \E_S\left( \langle \phi_k, K(S \mathbf{h}_j,\cdot)\rangle^2\right)\\
&=\varepsilon^{-1}\frac{1}{n}\sum_{j=1}^n \E_S\Big(\left\| P_{V_{i-1}^\perp} K(S \mathbf{h}_j,\cdot)\right\|_\mathcal{H}^2 \Big),
\end{aligned}$$
where $P_{V_{i-1}^\perp}$ denotes the orthogonal projection onto ${\rm span}\{\phi_k\}_{k=i}^\infty$.
The continuity of $K(\cdot,\cdot)$ implies that $f(\mathbf{h}) :=\E_S\Big(\left\| P_{V_{i-1}^\perp} K(S \mathbf{h},\cdot)\right\|_\mathcal{H}^2 \Big)$ is a continuous function in $\mathbf{h}$. Therefore, it follows from $\frac{1}{n}\sum_{i=1}^n \delta_{\mathbf{h}_i}\overset{d}{\longrightarrow}\nu$ that
{\scriptsize
$$\lim_{n\to\infty} \frac{1}{n}\sum_{j=1}^n \E_S\Big(\Big\| P_{V_{i-1}^\perp} K(S \mathbf{h}_j,\cdot)\Big\|_\mathcal{H}^2 \Big) = \E_{S,\mathbf{H}}\Big(\Big\| P_{V_{i-1}^\perp} K(S \mathbf{H},\cdot)\Big\|_\mathcal{H}^2 \Big). $$
}
Therefore,
$$\begin{aligned}
&\lim_{i\to\infty}\limsup_{n\to\infty} \mathbb{P}\Big( \sum_{k=i}^\infty S_n(\phi_k)^2\geq \varepsilon\Big) \\
&\leq \lim_{i\to\infty}\E_{S,\mathbf{H}}\Big(\left\| P_{V_{i-1}^\perp} K(S \mathbf{H},\cdot)\right\|_\mathcal{H}^2 \Big) = 0.
\end{aligned}$$
The last equality follows from the dominated convergence theorem, since
$\Big\| P_{V_{i-1}^\perp} K(S \mathbf{H},\cdot)\Big\|_\mathcal{H}^2 \leq K(S \mathbf{H},S \mathbf{H})$ is bounded.

Hence, all three conditions of Theorem~\ref{thm:kernel_CLT_tool} are verified, and thus the CLT is established. \qed

\section{Finite Sample Performance}\label{sec:OT-MMD-simu}
In this section, we examine the finite sample behavior of the proposed method in Section~\ref{sec:one-sample-kernel-OT} and compare it with existing procedures.
We consider testing two types of symmetry --- {\it spherical symmetry} and {\it exchangeability}. We set the symmetrized reference distribution $\nu_S$ as $\mathcal{N}(\mathbf{0},\mathbf{I}_p)$, as suggested in \cite[Section 4.2]{huang2023multivariate}. We use the Gaussian kernel $K(\mathbf{x},\mathbf{y}) :=\exp\left(-\frac{\|\mathbf{x}-\mathbf{y}\|^2}{4p}\right)$ on $\mathbb{R}^p$ --- this choice of bandwidth approximately fits to the scale of the distance between two observations from $\nu_S$, since under $\G$-symmetry ${\rm H}_{\G,0}$, the signed-ranks follow $\nu_S$, so $\mathbb{E}_{\mathbf{x}, \mathbf{y}\sim \nu_S}\frac{\|\mathbf{x}-\mathbf{y}\|^2}{4p}$ is a constant free of $p$. The level of the tests are set to 0.05, and the empirical power is obtained from 1000 replications.  Within each replication, the empirical reference distribution $\nu_n$ is obtained by drawing $n$ i.i.d.\ observations from $\nu$ and fixing it. With a fixed $\nu_n$, we can generate signed-ranks from the distribution-free null distribution. We generate 1000 such $T_n$'s from the null distribution and reject ${\rm H}_{\G,0}$ if the observed $T_n$ is greater than 95\% of the generated $T_n$'s. \newline

\noindent{\bf Spherical Symmetry}:
We first consider testing spherical symmetry, i.e., $\G={\rm O}(p)$. We use the same $\nu$ as in \cite[Remark 3.3]{huang2023multivariate}, which is the distribution of $\left(\sqrt{\chi^2_p},0,\ldots,0\right)$, where $\chi^2_p$ denotes a chi-square random variable with $p$ degrees of freedom. This reference distribution corresponds to a Gaussian effective reference distribution.
The following examples are from \cite[Section 4]{huang2023multivariate}:
\begin{enumerate}
\item[Sp1.] Multivariate Gaussian: $\mathbf{X}\sim \mathcal{N}\left(\mathbf{0}, \mathbf{I}_2\right) + \lambda\cdot \mathbf{1}_2$; $n=200$.
\item[Sp2.] Multivariate $t$-distribution with 1 degree of freedom, location parameter $\lambda\cdot \mathbf{1}_2$ and scale parameter $\mathbf{I}_2$; $n=200$.
\item[Sp3.] Uniform distribution over the unit disk $\{\mathbf{x}\in\R^2 : \|\mathbf{x}\|\leq 1\}$, plus $\lambda\cdot \mathbf{1}_2$; $n=200$.
\item[Sp4.] Elliptical: $n=200$, $p=2$, $\mathbf{X}=(2Z_1,Z_2)^\top$, where $Z_1$, $Z_2$ i.i.d.\ $\mathcal{N}(0,1)$.
\item[Sp5.] Correlation: $n=200$, $p=2$, $\mathbf{X}\sim \mathcal{N}\left(\mathbf{0},
\begin{pmatrix}
1&\rho\\
\rho&1\\
\end{pmatrix}\right)
$, where $\rho=0.6$.
\item[Sp6.] Chi-squared distribution: $n=100$, $p=2$, $X_1$, $X_2$ i.i.d.\ following $\chi^2(1)-1$.
\item[Sp7.] High-dimensional (${\rm H}_{\G,0}$): $n=200$, $p=50$, $\mathbf{X}\sim \mathcal{N}(\mathbf{0},\mathbf{I}_p)$.
\item[Sp8.] High-dimensional (${\rm H}_{\G,1}$): $n=200$, $p=50$, $\mathbf{X}\sim \mathcal{N}(\mathbf{0},\mathbf{I}_p) + 0.05\cdot \mathbf{1}_p$.
\item[Sp9.] High-dimensional (${\rm H}_{\G,1}$, heavy-tailed): $n=200$, $p=50$, $\mathbf{X}$ follows the multivariate $t$-distribution with 1 degree of freedom, location parameter $0.05\cdot \mathbf{1}_p$ and scale parameter $\mathbf{I}_p$.
\end{enumerate}
The generalized Wilcoxon's signed-rank test and generalized sign test \citep[Section 4]{huang2023multivariate} are denoted by ``OT-Wilcox'' and ``OT-sign'', respectively. It has been shown that OT-Wilcox is no worse than the Hotelling's $T^2$ test (denoted by ``$T^2$'') in detecting location shift alternatives (in terms of ARE). 
We also compare with \cite{Baringhaus1991} (denoted by ``LB''), a test for multivariate spherical symmetry that is consistent against any fixed alternative.
A function $h$ needs to be specified for LB. When $p=2$, we use the function $h(t)=\frac{t-1/4}{17/8-t}$ as in \cite[Section 4]{Baringhaus1991}, which yields a tractable asymptotic null distribution.
When $p=50$, we use the function $h(t)=(1-2tw+w^2)^{-\lambda}-1$ \citep[Equation 3.10]{Baringhaus1991} with $\lambda=24$ and $w=1/4$ as in \cite[Section 5]{Baringhaus1991}, and
the same first-order approximation to the upper tail probability for computing the critical value, as advocated in \cite[Equation 4.4]{Baringhaus1991}.
We also compare our methods with \cite{Henze2014sph} (denoted by ``HHM''), a test for spherical symmetry based on the empirical characteristic function. We use the suggested settings in \cite[Section 5]{Henze2014sph}.\footnote{More specifically, we use the Kolmogorov-Smirnov type statistic with the hyper-parameter $\mathcal{R}=2$. For $p=2$, we use 8 rings and 9 grid points as in \cite[Section 5]{Henze2014sph}. For $p=50$, we use 8 rings and 200 grid points).}


Table~\ref{tab:sph} presents the empirical power of various methods for testing spherical symmetry. Our proposed test, denoted by ``OT-MMD'', stands out as one of the best-performing methods across different scenarios. It demonstrates strong power not only against location-shift alternatives (Sp1–Sp3) but also against other alternatives (Sp4–Sp9). Even in the Gaussian location model (Sp1), where OT-Wilcox and $T^2$ are theoretically optimal \citep[Theorem 3.9]{huang2023multivariate}, OT-MMD performs comparably. Moreover, in non-location-shift models (Sp4–Sp6), where OT-Wilcox and $T^2$ become powerless, OT-MMD remains powerful. Notably, in the high-dimensional settings (Sp7–Sp9), where LB and HHM are powerless, OT-MMD maintains high power. \newline

\begin{table*}
\centering
\caption{Power of the competing methods for testing spherical symmetry, with nominal level 0.05.}
\label{tab:sph}
\begin{tabular}{c||cccccc}
 \toprule
 \multicolumn{7}{c}{Multivariate Gaussian Distribution} \\
 \hline
Sp1& $T^2$ & OT-Wilcox & OT-sign &LB & HHM&OT-MMD\\
 \hline
$\lambda = 0.00$ & 0.06  & 0.05 & 0.05 & 0.05&0.05 &0.05\\
$\lambda = 0.05$ & 0.15  & 0.14 & 0.07 & 0.08& 0.10 &0.13\\
$\lambda = 0.10$ & 0.42  & 0.42 & 0.12 & 0.18&0.26 &0.40\\
$\lambda = 0.15$ & 0.77  & 0.76 & 0.21& 0.38&0.54 &0.75\\
$\lambda = 0.20$ & 0.95  & 0.95& 0.36 & 0.62& 0.84&0.95\\
\midrule
 \multicolumn{7}{c}{Multivariate $t$ Distribution} \\
 \hline
Sp2& $T^2$ & OT-Wilcox & OT-sign &LB & HHM&OT-MMD\\
 \hline
$\lambda = 0.0$ & 0.03  & 0.07& 0.05 & 0.04& 0.05&0.06\\
$\lambda = 0.1$ & 0.05  & 0.16& 0.09 & 0.21& 0.19&0.21\\
$\lambda = 0.2$ & 0.06  & 0.46 & 0.20 & 0.75& 0.60&0.67\\
$\lambda = 0.3$ & 0.07  & 0.79 & 0.39 & 0.97& 0.94&0.93\\
$\lambda = 0.4$ & 0.10  & 0.96 & 0.62 & 1.00& 0.99&1.00\\
\midrule
\multicolumn{7}{c}{Multivariate Uniform Distribution} \\
 \hline
Sp3& $T^2$ & OT-Wilcox & OT-sign &LB & HHM&OT-MMD\\
 \hline
$\lambda = .000$ & 0.06  & 0.05 & 0.05 & 0.04& 0.05&0.03\\
$\lambda = .025$ & 0.14  & 0.21 & 0.06 & 0.04& 0.09&0.08\\
$\lambda = .050$ & 0.40  & 0.61& 0.09 & 0.05& 0.25&0.36\\
$\lambda = .075$ & 0.75  & 0.90 & 0.14 & 0.07& 0.55&0.69\\
$\lambda = .100$ & 0.96  & 0.99 & 0.23 & 0.11& 0.83&0.91\\
\midrule
\multicolumn{7}{c}{Other Alternatives} \\
 \hline 
Sp4-Sp6 & $T^2$ & OT-Wilcox & OT-sign &LB & HHM&OT-MMD\\
\hline 
Elliptical & 0.06 & 0.06 &0.05 &0.50 & 1.00&1.00\\
Correlation &0.06 &0.06  &0.05  &0.50 & 1.00&1.00\\
Chi-squared & 0.09 & 0.27  &0.45  &1.00 & 1.00&0.88\\
\midrule
\multicolumn{7}{c}{High-Dimensional Settings} \\
\hline
Sp7-Sp9 & $T^2$ & OT-Wilcox & OT-sign &LB & HHM&OT-MMD\\
\hline
High-dim (${\rm H}_{\G,0}$) &0.04 &0.05  &0.05 &0.00 &0.04&0.05\\
High-dim (${\rm H}_{\G,1}$) &0.54 &0.68 &0.05 &0.00 &0.15&0.68\\
High-dim (${\rm H}_{\G,1}$, heavy-tailed) & 0.05 & 0.35 & 0.05 & 0.00 & 0.11&0.39\\
\bottomrule
\end{tabular}
\end{table*}

\noindent{\bf Exchangeability}:
Next, we consider the problem of testing exchangeability. OT-Wilcox is not applicable in this setting. We set $\nu_S = \mathcal{N}(\mathbf{0}, \mathbf{I}_p)$. Let $q(\mathbf{x}) :=(x_{(1)},\ldots,x_{(p)})$ be the reordering of $\mathbf{x} \in \R^p$ from the smallest to the largest, and let $\nu = q\# \nu_S$.

For comparison, we include the bivariate exchangeability tests proposed by \cite{EG2016} (denoted by EG1 and EG2) and \cite{modarres2008tests} (denoted by R, T3, M, S, D). Following~\cite{EG2016}, we consider the following scenarios:
\begin{enumerate}
\item ${\rm H}_0^a$: $X,Y\sim \mathcal{N}(0,1)$
\item ${\rm H}_1^a$: $X\sim \mathcal{N}(2,1)$, $Y\sim \mathcal{N}(0,1)$
\item ${\rm H}_1^b$: $X\sim \mathcal{N}(1,1.5^2)$, $Y\sim \mathcal{N}(0,1)$
\item ${\rm H}_1^c$: $X\sim {\rm Exp}(1)$, $Y\sim \mathcal{N}(0,1)$
\item ${\rm H}_1^d$: $X\sim \mathcal{N}(0,1)$, $Y\sim {\rm LogNormal}(0,1)$
\end{enumerate}

As observed in the case of testing spherical symmetry, Table~\ref{tab:exchange} demonstrates that OT-MMD exhibits high power against various alternatives (${\rm H}_1^a$ - ${\rm H}_1^d$). Across all settings, our method consistently achieves the best or nearly best performance in terms of power. \newline

\begin{table*}
\centering
\caption{Power of the competing methods for testing exchangeability, with nominal level 0.05.} 
\label{tab:exchange}
\begin{tabular}{l||cccccccc}
 \toprule
 \multicolumn{9}{c}{$n=50$} \\
 \hline
 & EG1 & EG2 & R & T3 & M &S & D & OT-MMD\\
 \hline
${\rm H}_0^a$ & 0.04  & 0.03 & 0.05 & 0.04&0.05 & 0.04&0.05 & 0.05 \\
${\rm H}_1^a$ & 1.00  & 1.00 & 0.43 & 0.33&0.23 & 1.00&1.00 &1.00 \\
${\rm H}_1^b$ & 0.69  & 0.39 & 0.36 & 0.26&0.22 & 0.80&0.77 & 0.71\\
${\rm H}_1^c$ & 0.98  & 0.95 & 0.50 & 0.52&0.60 & 0.18&0.36 & 0.92\\
${\rm H}_1^d$ & 1.00  & 0.99 & 0.54 & 0.59&0.89 & 0.44&0.58 & 0.99\\
\midrule
 \multicolumn{9}{c}{$n=100$} \\
 \hline
& EG1 & EG2 & R & T3 & M &S & D & OT-MMD\\
 \hline
${\rm H}_0^a$ & 0.05  & 0.04 & 0.05 & 0.05&0.05 & 0.05&0.05 & 0.05 \\
${\rm H}_1^a$ & 1.00  & 1.00 & 0.58 & 0.49&0.28 & 1.00&1.00 &1.00 \\
${\rm H}_1^b$ & 0.95  & 0.81 & 0.39 & 0.54&0.34 & 0.99&0.95 &0.97 \\
${\rm H}_1^c$ & 1.00  & 1.00 & 0.77 & 0.79 &0.94 & 0.37 & 0.64 &1.00 \\
${\rm H}_1^d$ & 1.00  & 1.00 & 0.81 & 0.89&1.00 & 0.78&0.94 &1.00 \\
\bottomrule
\end{tabular}
\end{table*}

\noindent{\bf Real Data Example}:
In analyzing the returns of multiple financial assets over time, testing for exchangeability helps determine whether the returns can be considered as coming from a common distribution. This, in turn, can inform decisions about grouping assets for downstream analysis and applications. For example, when multiple assets are exchangeable, semi-static hedging techniques can be employed for multi-asset barrier options \citep{molchanov2011exchangeability}.


We examine the monthly returns of equities from various companies over a five-year period, spanning from January 1, 2019, to December 1, 2023. Let $P_t$ denote the adjusted\footnote{The {\it adjusted close price} accounts for stock splits, dividends, and capital gain distributions. The data used in this analysis are sourced from \url{https://finance.yahoo.com}.} close price of an asset on the first trading day of the $t$-th month. The monthly return is then computed as $R_t = \frac{P_t}{P_{t-1}} - 1$.


We first analyze the monthly returns of three technology companies: NVIDIA Corporation, Apple Inc., and Tesla, Inc., testing whether their returns are exchangeable. Using the same setting and choice of $\nu$ as above, we obtain a $p$-value of 0.87, suggesting no strong evidence against exchangeability. This result may stem from the fact that all three companies belong to the technology sector and exhibit similar return behaviors.

However, when replacing Tesla with the healthcare company Pfizer Inc., the exchangeability test for NVIDIA Corporation, Apple Inc., and Pfizer Inc. yields a highly significant $p$-value $<0.001$. This indicates that companies from different sectors may have fundamentally different return dynamics. The factors influencing the healthcare company Pfizer's returns likely differ substantially from those driving the returns of technology companies.
\end{appendix}

\bibliographystyle{imsart-number} 
\bibliography{symmetry}       


\end{document}